**Title: The Role of Affect and Priors in the Generation of Hallucinations in Early Psychosis**

Authors: Timothy Friesen[1], Philomène Labilloy[1], Deven Parekh[1], Maia List[1], Youcef Barkat[2], Victoria Fisher[3], Camille Jacquet[4], Jed Burde[5], Ivy Guo[1], Eva Hammer[1], Parnia Akhavansaffar[1], Dylan Hamitouche[1], Martin Lepage[1], Chris Mathys[6], Jai Shah[1], Al Powers[5], David Benrimoh[1]

1.  McGill University

2.  Université de Montréal

3.  Columbia University

4.  Université Grenoble Alpes

5.  Yale University

6.  Aarhus University

**Corresponding Author: Timothy Friesen**

**Address: 6875 Boulevard LaSalle, H4H 1R3, Verdun, Québec, Canada**

**University Affiliation: McGill University**

**Phone Number: 204-612-0249**

**Email: timothy.friesen@mail.mcgill.ca**

**Abstract:**

Background: Stress and negative affect play significant roles in developing psychosis. Bayesian analyses applied to the conditioned hallucinations (CH) task suggest that hallucinations arise when maladaptive prior beliefs outweigh sensory evidence. Prior weighting is linked to hallucination severity, yet the nature of these priors remains unclear. Negative affect may influence the strength of maladaptive priors. We hypothesized that, under stress, participants will show increased CH rates and prior weighting, with this effect more pronounced in patients.

Methods: This study employs a modified CH task using valenced linguistic stimuli and stress and non-stress affective manipulations. The sample for this pilot study included those at risk for psychosis and patients with first episode psychosis (N=12) and healthy controls (N=15). The objective of this study was first to validate this affective version of the CH task and then to demonstrate an effect of affect on CH rates and prior weighting.

Results: Replicating past results, patients had higher CH rates ($b$ = 0.061, $p$ < 0.001) and prior weighting ($b$ = 0.097, $p$ < 0.001) for session 1 compared to controls (n=15) across conditions. Further, runs with stress manipulations had higher prior weighting across patients and controls compared to runs with non-stress manipulations ($b$ = 0.054, $p$ = 0.033).

Conclusions: This study validates this affective version of the CH task and provides preliminary evidence of a relationship between affective state and prior weighting. Future work will be aimed at confirming and extending these findings, with the objective of developing biomarkers of early psychosis.

**Introduction**

There is an acute need to improve treatments for patients in the earliest phases of psychosis. Accomplishing this requires improving the measures available to characterize the latent processes which lead to psychosis onset and the development of psychotic symptoms (D. Benrimoh et al., 2025; D. Benrimoh, Fisher, Mourgues, et al., 2023; A. Powers et al., 2025). In recent work, we have argued that negative affect may play an important role in the development of psychotic symptoms (D. Benrimoh et al., 2025), shifting expectations towards negative valences, accounting for the preponderance of negatively valenced delusions and hallucinations in clinical psychosis. Indeed, it has long been known that affective changes occur early in the prodrome, the period leading up to psychosis onset, and that stress, especially social stress, and mood changes can affect psychotic symptoms (van Winkel et al., 2008). Depression and anxiety are both positively correlated with the distressing aspects of positive symptoms such as auditory verbal hallucinations (Hartley et al., 2013). Recent daily social stress has also been correlated with the frequency and distress of auditory hallucinations (Farina et al., 2024). In addition, in schizophrenia, anxiety has been shown to be positively correlated with positive symptoms: those with severe anxiety suffer from more hallucinations as well as increased social isolation and hopelessness compared to those with lower anxiety (Lysaker & Salyers, 2007). Experience Sampling Methods (ESM) assessing psychotic symptoms in patients suggests that negative affect precedes positive psychotic symptoms (Kramer et al., 2014).

The tools of computational psychiatry now allow us to estimate parameters from behavioral data, facilitating the interrogation of latent states underlying psychotic symptom development. Models rooted in Bayesian inference offer mechanistic explanations of how disruptions in internal beliefs and perceptual processes give rise to various mental health symptoms (D.

Benrimoh et al., 2018). Indeed, in both in silico and in vivo work we (D. Benrimoh et al., 2018)(D. Benrimoh et al., 2019; Kafadar et al., 2022; A. R. Powers et al., 2017) have demonstrated that hallucinations can occur when the strength of maladaptive prior beliefs outweighs reliance on sensory evidence. However, the specific nature of the priors underlying psychotic symptoms, including their valence and their sensitivity to stress, remains unclear (D. A. Benrimoh & Friston, 2020)(D. Benrimoh et al., 2022)(D. A. Benrimoh & Friston, 2020)(Sterzer et al., 2018)(D. A. Benrimoh & Friston, 2020).

The conditioned hallucination (CH) task has consistently demonstrated that higher prior weighting correlates with hallucinations and symptom severity (D. Benrimoh et al., 2024; Kafadar et al., 2022; A. R. Powers et al., 2017). We propose that negative affective states may provide important context for the development, maintenance, and strength of maladaptive priors. In this work, we set out to design and perform the initial validation of an affective version of the CH task (Aff-CH). The first objective of this pilot study was to ensure that these modifications to the CH task still produced the same findings as earlier iterations of the CH task - namely, increased CH rates and prior weighting in patients compared to controls. We further hypothesized that negatively valenced priors and stress manipulations would generate higher CH rates and prior weighting; that this effect would be strongest in patients; and that congruence between the valence of the manipulation and prior would show stronger CH rates and prior weighting than incongruent conditions.

**Methods**

**Sample:**

Participants were invited to complete the Aff-CH task at two sessions, separated by at least one month. We included two sessions to examine the test-retest reliability for response and CH rates. Due to sample size limitations in the second session, we focused on differences at session 1, and analyses averaged across both sessions. Patients were recruited from a first episode psychosis program and from a clinic for those deemed at clinical high risk of developing psychosis (CHR-P). Due to sample size limitations, these two populations were grouped together as our early psychosis patient group. While patients with CHR-P do not have diagnosed psychosis and may therefore have differing results, we note that in previous work, we have observed similar results in patients with psychosis as well as those with high psychosis proneness in the general population, with the task more sensitive to hallucination status than diagnosis per se(D. Benrimoh, Fisher, Seabury, et al., 2023; A. R. Powers et al., 2017). Both clinics are located in Montreal, Canada and provide services under public healthcare and serve a diverse population. Controls were 18 and older, had no current or past psychotic illness, and had had no active mental health followup for non-psychotic reasons for at least six months. This definition of healthy control was chosen in order to make these controls a more realistic sample of the general population (rather than controls who have never experienced mental illness), from which patients at risk for psychosis must be differentiated. Patients needed to be 15-34 years old (consistent with clinic admission criteria), and must have experienced, at some time in the last three months, auditory hallucinations or altered auditory perceptual experiences. They could not have a diagnosis of purely substance-induced psychosis, or cognitive impairment precluding participation in the study.

**Affective CH Task:**

*Rationale*

In the CH task (see detailed description in (A. R. Powers et al., 2017)), a conditioned association is established between a neutral visual stimulus and a short (~1s) auditory stimulus, in the presence of white noise. A QUEST procedure is used first to establish audibility thresholds. After this, whenever the visual stimulus is presented, the auditory stimulus is presented at volumes estimated to result in 75%, 50%, 25%, or 0% reported detection rates, based on the threshold determined during QUEST and a standard psychometric curve. Over the course of the experiment, the proportion of lower-audibility trials increases. Participants are asked, after each presentation of the visual stimulus, to indicate via button press if they believe they heard the auditory stimulus or not; and then to rate their confidence in that judgement by holding down a button for a longer or shorter amount of time. When a participant indicates hearing the auditory stimulus on a 0% trial, this is considered a conditioned hallucination. We retained this basic structure in the current experiment; however we utilized a shortened version of the task lasting roughly 15 minutes per run after previous pilot work identified this length as being viable. As described below, this was necessary to maintain feasibility given the need to have participants complete multiple runs in a single session.

Our objective was to validate a version of the CH task where affect is modulated to examine the effect of affect in the generation of hallucinations in early psychosis. We wanted to determine the impact of the affective state being manipulated *after* the prior was instilled, in order to simulate the impact of stress on affective state and, subsequently, the valenced priors underlying psychotic symptoms. As such we modified the CH task in two key ways.

First, for the auditory stimulus, we replaced the tone with a positively or negatively valenced word. Next, we added a valenced (stress or non-stress) manipulation once the prior was established (which previous piloting suggested occurred after the first block of 45 trials). After the manipulation, the task continued for two blocks, a total of 90 trials. The purpose of this manipulation was to modulate the strength of the prior. By introducing a manipulation after the prior was established, a significant concern was an accelerated decrease in the prior strength, leading to lower CH rates and prior weighting after the manipulation; as such, ensuring this did not occur was an important part of task validation.

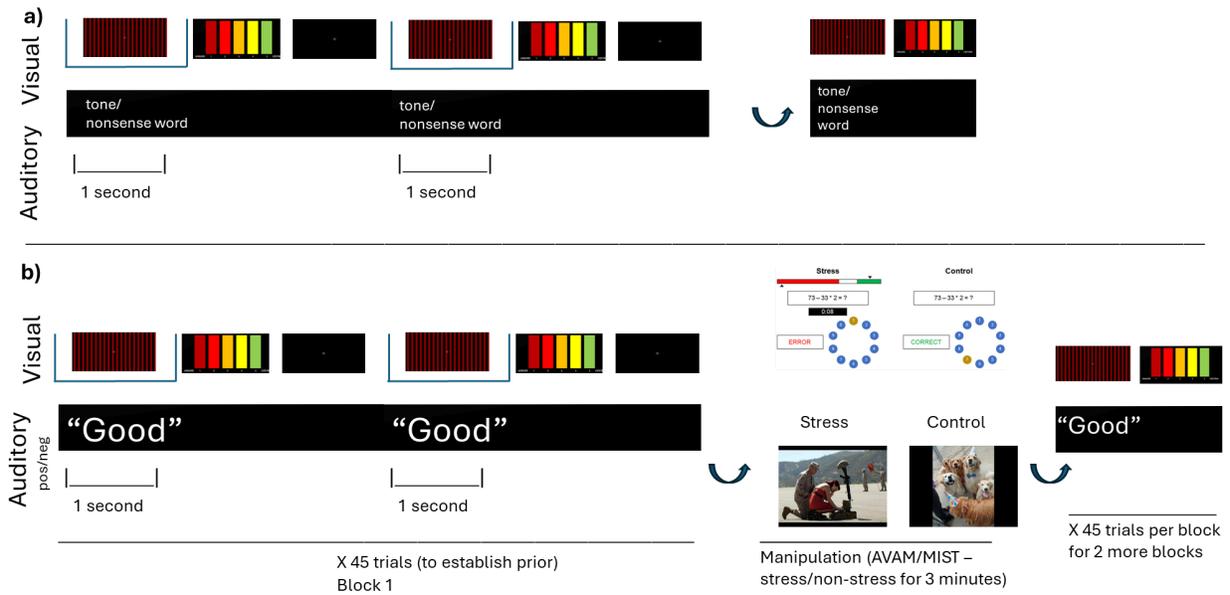

**Figure 1: Description of the Affective CH task**

Run structure of the original and affective version of the CH task. Panel (a) shows the structure of the original CH task where after completing test trials and the QUEST procedure, participants are presented with a word (Powers *et al.*, 2017) or a nonsense word (Benrimoh *et al.*, 2024) in conjunction with a visual stimulus and participants must indicate if they heard the word or not, followed by how confident they are. In panel (b), the overall structure of the task remains the same, but the word is replaced with a valenced word. Participants complete 3 blocks of 45 trials. Blocks 1 and 2 are separated by a manipulation (either AVAM or MIST depending on the session) that is either of stress or non-stress valence and participants complete blocks 2 and 3 after the valenced manipulation.

*Auditory Stimuli*

The auditory stimulus was an affectively laden word (e.g. 'joy', 'hate') selected from a database of words with known affective valences (see Supplementary Materials for the full list of words used) (Warriner et al., 2013). Subjects were randomized to a positive word or a negative word for each run. The words presented were lengthened using Audacity software in order to last roughly 1 second, consistent with (D. Benrimoh et al., 2024). See supplementary materials for validation analyses.

*Visual Stimuli*

The visual stimuli were a set of non-valenced colored patterned images (e.g. a screen with green vertical lines). See figure 1 for an example of visual stimuli.

*Affective Manipulations*

We piloted two affective manipulations, the existing Montreal Imaging Stress Task (MIST) and an Audiovisual Affective Manipulation (AVAM), which was created for this study. Each manipulation had a stress and non-stress condition. For these pilot analyses we consider these manipulations together; in the supplementary, we present results subgrouped by manipulation type. We grouped the manipulations together as MIST has been shown to reliably produce stress (Dedovic et al., 2005) and we validated that the AVAM can modulate the affective state (see the supplementary materials for this validation and disaggregated results).

*Montreal Imaging Stress Task (MIST)*

The MIST (Dedovic et al., 2005) is a stress-induction task in which the subject performs mental math and experiences social evaluation of their skills. The MIST was randomly set to the non-stress (no time limit to respond to questions; social evaluation indicating adequate performance) or stress condition (questions more difficult than the patient is capable of completing; negative social evaluation). The social evaluations were pre-recorded audio files based on the script from the original MIST experiment. For the non-stress social evaluation, the audio recording involved telling the participant that they are performing as expected and to keep up their performance. For the negative social evaluation, the audio recording involved telling the participant that they are performing poorly on the task and if their poor performance continues, they may need to be removed from the experiment. The MIST lasted roughly three minutes.

*Audiovisual Affective Manipulation (AVAM)*

The AVAM was a 3-minute audio-video presentation of either positive images and music (non-stress condition) or stressful, aversive images and music (stress condition). All music was instrumental and did not involve language. Music was chosen as in initial pilot testing it generated a stronger subjective emotional reaction than other sound effects. Images and audio were drawn from existing validated research stimulus repositories (International Affective Picture System - IAPS and Open Affective Standardized Image Set - OASIS) (Lang, P.J.,

Bradley, M.M., & Cuthbert, B.N., 1997), (Kurdi et al., 2017). The AVAMs were validated in a group of healthy controls (see the supplementary materials).

*Overall task design*

At each of two sessions, one month apart, participants were invited to complete runs with different manipulation of word valence combinations (stress manipulation - negative word valence, non-stress manipulation - negative word valence, stress-manipulation - positive word valence, non-stress manipulation - positive word valence). As not all participants completed both sessions, for these pilot analyses we report results averaged across the combined sessions, unless otherwise noted.

**Measures**

*Measures - affect*

We used two measurements of affect throughout the experiment: an affective slider (Betella & Verschure, 2016) which was completed four times per run (at the beginning of the run, before the manipulation, after the manipulation, and at the end of the run) to measure arousal and valence (see supplementary materials for the visual shown to participants); and the short version of the Immediate Mood Scalar (IMS)(Nahum et al., 2017) at the beginning and end of each run (see supplementary materials for the list of IMS items).

*Clinical measures*

We used the Patient Health Questionnaire (PHQ-9) (Kroenke et al., 2001), Generalized Anxiety Disorder (GAD-7) (Spitzer et al., 2006), Launay-Slade Hallucinations Scale (LSHS) (Launay & Slade, 2012), and Auditory Hallucinations Rating Scale (AHRS) (Hoffman et al., 2021) to assess baseline clinical symptoms before every session.

*Computational modeling - Hierarchical Gaussian Filter (HGF)*

Using the Hierarchical Gaussian Filter, we estimated three levels of perception where X1 was defined as whether or not the participants heard the word and X2 describes the belief in the association between the word and visual stimuli. We estimated ν (nu), which is embedded in the X2 term on the second level, to estimate the precision of the prior. Our main computational parameter of interest is the prior weighting parameter (ν), which represents the relative weighting of prior beliefs and sensory evidence. Higher ν values represent a higher weighting of prior beliefs compared to incoming sensory information. For our documentation relating to the HGF model we used in Julia, please visit our github (https://github.com/McPsyt/HGF-Analysis-Julia).

*Statistical Analysis*

Linear mixed models, accounted for repeated observations per participant, were used to test main effects of group (patient vs. controls) and manipulation valence (stress vs. non-stress) on

CH rates, v and confidence measures. Spearman correlation analysis was used to correlate CH rate and v and symptom measures with CH rate and v. Mann-Whitney U tests were used to compare patient and control groups on clinical measures. As this was a pilot analysis, we did not perform correction for multiple comparisons. Test-retest correlations were examined for the CH rates and v values across sessions 1 and 2. We used an intraclass correlation coefficient (ICC) function and selected the one-way random effects ICC using the single measure level, which gives the reliability of a single rater's score (each session or run). All analyses were carried out in python (Version 3.12.10). As this was an exploratory pilot analysis, analyses were not corrected for multiple comparisons.

## Results

*Sample*

We recruited 16 patients that were either FEP (n = 9) or CHR-P (n = 7) and 15 healthy controls (see Table 1). 16 patients attended the first session, of whom 12 patients had useable data (see below); of these, 7 attended the second session. 15 healthy controls completed the first session and 13 healthy controls completed the second session.

Four patients were excluded from analysis: two were clinically unstable at the time of testing and were deemed unable to complete the experiment; one patient indicated reactivation of symptoms during the task and left the experiment early; and one patient was later deemed by their clinician to have a purely drug-induced psychosis. The final analysis sample included 12 patients (5 FEP and 7 CHR-P). See Table 1 for demographics of the analysis sample.

Mean LSHS scores at baseline were significantly higher in the patient compared to control groups ($U$ = 13.5, $p$ < 0.001). Mean PHQ-9 scores at baseline were significantly higher in the patient compared to control groups ($U$ = 32.0, $p$ = 0.001). Mean GAD-7 scores at baseline were significantly higher in the patient compared to control groups ($U$ = 33.0, $p$ = 0.002). Within the patient group, patients with FEP compared to CHR-P patients had lower LSHS, PHQ-9, and GAD-7 scores although none of these differences (LSHS: $U$ = 15.5, $p$ = 0.807, PHQ-9: $U$ = 9.0, $p$ = 0.192; GAD-7: $U$ = 11.5, $p$ = 0.369) were significant. See Table 1 for a full list of symptom scores by group. Table S1 for a list of symptom scores and medication status split by FEP and CHR-P patients. Of the FEP patients, 100% (5/5) were taking antipsychotics during the first session and 100% (2/2) were taking antipsychotics during the second session. Of the CHR-P patients, 43% (3/7) were on antipsychotics during the first session and 50% (3/6) were on antipsychotics for the second session.

**Table 1: Demographics**

| Group | All | Patient | Control |
|---|---|---|---|
| Age | Mean: 23.26 | Mean: 22.75 | Mean: 23.47 |

|  |  | Std: 5.66 | Std: 5.33 | Std: 6.64 |
|---|---|---|---|---|
| Sex |  | Female: 20<br>Male: 7 | Female: 8<br>Male: 4 | Female: 12<br>Male: 3 |
| EthnicityF |  | European origins: 55.6%<br>Middle Eastern origins: 18.5%<br>Asian origins: 18.5%<br>African origins: 11.1%<br>Caribbean origins: 7.4%<br>Hispanic origins: 3.7% | European origins: 41.7%<br>Middle Eastern origins: 16.7%<br>Asian origins: 25%<br>African origins: 16.7%<br>Caribbean origins: 16.7%<br>Hispanic origins: 0% | European origins: 66.7%<br>Middle Eastern origins: 20%<br>Asian origins: 13.3%<br>African origins: 6.7%<br>Caribbean origins: 0%<br>Hispanic origins: 6.7% |
| Mean LSHS scores |  | Mean: 13.259<br>Std: 13.375 | Mean: 25.667<br>Std: 10.094 | Mean: 3.333<br>Std: 3.830 |
| Mean PHQ-9 scores |  | Mean: 6.444<br>Std: 6.034 | Mean: 10.125<br>Std: 6.355 | Mean: 3.200<br>Std: 3.764 |
| Mean GAD-7 scores |  | Mean: 7.111<br>Std: 5.820 | Mean: 10.917<br>Std: 5.632 | Mean: 4.067<br>Std: 3.955 |
| Diagnosis |  |  | FEP: 5 (41.7%)<br>CHR: 7 (58.3%) |  |

### *The Aff-CH task replicates results from the original CH task*

Replicating past results (A. R. Powers et al., 2017)(D. Benrimoh et al., 2024; Kafadar et al., 2022)(A. R. Powers et al., 2017), we saw higher CH rates (Patient mean: 0.187, SD = 0.161; Control mean: 0.120, SD = 0.088) and higher *v* values (Patient mean: 0.655, SD = 0.159; Control mean: 0.560, SD = 0.141)  in the patient compared to the control group. The difference in CH rates between patients and controls was not significant when averaging across both sessions ($b = 0.067$, $SE = 0.048$, $z = 1.382$, $p = 0.167$), but was significant in session 1 ($b = 0.061$, $SE = 0.016$, $z = 3.764$, $p < 0.001$). For *v* values, the difference between patients and controls was significant when averaging across both sessions ($b = 0.095$, $SE = 0.043$, $z = 2.206$, $p = 0.027$) and in session 1 ($b = 0.097$, $SE = 0.004$, $z = 23.679$, $p < 0.001$). Further, using a Spearman correlation, we found that CH rates and *v* values were significantly correlated across all participants and both sessions ($ρ = 0.371$, $p = 0.013$). We found that the correlation between CH rates and *v* values was driven by the patient group who showed a stronger positive correlation ($ρ = 0.545$, $p = 0.016$) where the control group showed a weaker, non-significant, positive correlation ($ρ = 0.175$, $p = 0.403$). Looking at the range of CH rates and v values for patients and controls grouped across sessions and runs, we see that patients have higher minimum and maximum CH rates and v values compared to controls (patient min CH rate:

0.047, patient max CH rate: 0.525, HC min CH rate: 0.015, HC max CH rate: 0.318; patient min v value: 0.363, patient max v value: 0.817, HC min v value: 0.310, HC max v value: 0.757). Furthermore, as in (A. R. Powers et al., 2017), we found that as the audibility increased, the response rates of the patient group approached the values of the control groups, as shown in Figure 2.c. We found a significant main effect of group on response rates in the 25% ($b$ = -0.109, $SE$ = 0.051, $z$ = -2.140, $p$ = 0.032) and 50% audibility conditions ($b$ = -0.042, $SE$ = 0.002, $z$ = -25.286, $p$ < 0.001) where patients showed higher response rates compared to controls, but not in the 75% audibility condition ($b$ = 0.019, $SE$ = 0.048, $z$ = 0.392, $p$ = 0.695) where controls showed numerically higher response rates, but no significant group differences were found. For further replication of the original CH task, please see supplementary materials.

*Cumulative hallucination rates:* Further replicating past results (D. Benrimoh et al., 2024), we plotted cumulative CH rates as a function of trials and we see that patients have higher mean cumulative CH rates over all runs compared to controls (Cumulative CH rates - patient mean: 0.207, control mean: 0.127), however, this group difference was not significant ($b$ = 0.076, $SE$ = 0.051, $z$ = 1.489, $p$ = 0.137; Fig 2.d). Importantly, we do not see any disruption in the smooth decay of CH rates after the affective manipulation, which is crucial for demonstrating that the addition of this manipulation did not significantly alter task dynamics.

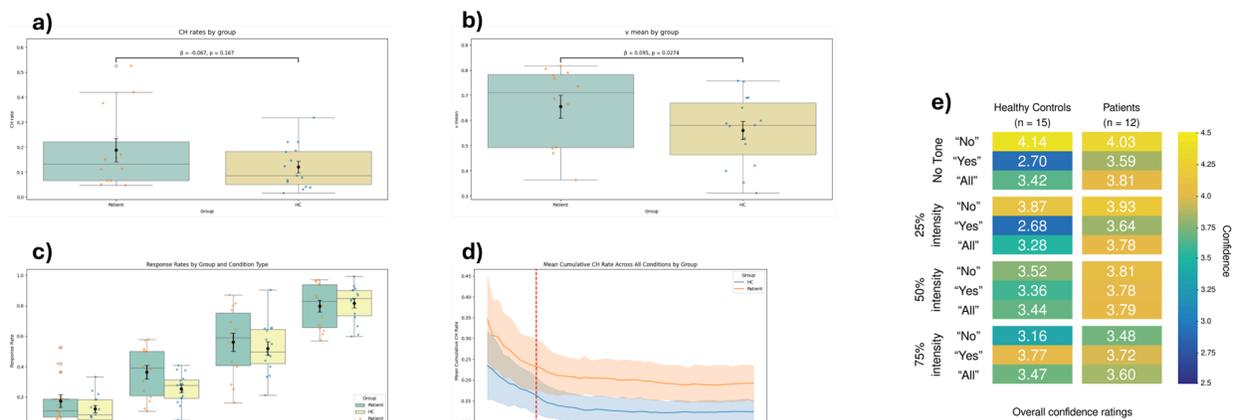

**Figure 2: Validation of original Affective CH experiment**

Panel (a) shows CH rates by group where boxplots display group medians with 95 % confidence intervals for the median. Overlaid black points represent group means ± standard error of the mean (SEM), while individual points correspond to each participant's average CH rate across sessions (or only session 1 if session 2 was not completed). Panel (b) shows v values by group where boxplots depict group medians with 95 % CIs for the median. Black points with vertical bars denote the group mean ± SEM, providing a parametric summary of perceptual precision estimates derived from the Hierarchical Gaussian Filter (HGF) model. Panel (c) shows response rates by trial type and group where boxplots represent the median proportion of "yes" responses across word-intensity levels (0 %, 25 %, 50 %, 75 %) for each group, with boxes indicating 95 % CIs for the median and black points showing mean ± SEM. Individual points reflect participant-level averages, illustrating group differences in detection behavior across increasing stimulus intensities. Panel (d) shows the mean cumulative proportion of "yes" responses in no-word trials plotted over trial number for healthy controls (HC) and patients, with shaded bands representing 95 % confidence intervals around the mean. The vertical red dashed line at trial 10 marks the point at which the manipulation occurred. Panel (e) shows confidence ratings by group, wordpresence, and intensity. Heatmap showing mean confidence ratings (color-coded by magnitude) for "Yes," "No," and "All" responses across word-intensity levels (0 %, 25 %, 50 %, 75 %) and groups.

### *The impact of affect on prior weighting and conditioned hallucinations*

*Higher prior weighting in stress compared to non-stress manipulation runs*

Next, we examined the effect of the different valenced manipulations on CH rates and *v* by looking at CH rates and prior weighting across both groups for all non-stress runs versus stress runs, averaged across both sessions. Numerically, CH rates were higher in the stress compared to non-stress conditions (stress CH rate mean: 0.165, SD = 0.157; non-stress CH rate mean: 0.135, SD = 0.108) and we saw the same trend with v values (stress v mean: 0.628, SD = 0.157; non-stress v mean: 0.574, SD = 0.174). We looked at the difference between CH rates in non-stress and stress runs and did not observe a significant difference ($b$ = 0.031, $SE$ = 0.016, $z$ = 1.897, $p$ = 0.058). Importantly, however, stress runs had higher v values compared to non-stress runs ($b$ = 0.054, $SE$ = 0.025, $z$ = 2.130, $p$ = 0.033). This demonstrates that, across participants, the affective manipulation did influence prior weighting, in line with our hypothesis. When looking at measures from session 1, we did not find a significant main effect of manipulation valence on CH rates ($b$ = 0.020, $SE$ = 0.017, $z$ = 1.137, $p$ = 0.256), but this was significant for v values ($b$ = 0.076, $SE$ = 0.031, $z$ = 2.465, $p$ = 0.014). We also found a main effect of manipulation valence on cumulative CH rates (see Fig 3; $b$ = -0.030, $SE$ = 0.003, $z$ = -10.349, $p$ < 0.001), this effect remained significant when factoring in group and prior valence where cumulative CH rates were higher in the stress compared to the non-stress condition (stress cumulative CH rate: 0.176; non-stress cumulative CH rate: 0.147).

*Effect of manipulation valence between groups*

Further, we found a significant interaction between cumulative CH rates and group: cumulative CH rates were higher in the patient-stress conditions ($b$ = -0.035, $SE$ = 0.006, $z$ = -6.016, $p$ < 0.001). For v values, we did not find a significant group by manipulation valence interaction ($b$ = -0.017, $SE$ = 0.052, $z$ = -0.335, $p$ = 0.738). Numerically, the highest *v* and CH rate values were for patients in the stress manipulation condition. When looking at cumulative CH rates post manipulation, similar to CH rates, we found that patients had higher rates across all valence combinations, and we found a significant interaction between group, word, and manipulation valence ($b$ = 0.087, $SE$ = 0.012, $z$ = 7.454, $p$ < 0.001) where patients showed higher cumulative CH rates in the stress manipulation and negative word condition. This effect was driven by the group term as the interaction between manipulation and word valence was not significant ($b$ = -0.003, $SE$ = 0.008, $z$ = -0.442, $p$ = 0.659).

*Trend towards higher patient CH rates and v in the negative prior-stress manipulation condition*

We then looked at group differences for CH rates and v values for manipulation and word valence combinations. Most likely due to our small sample size for each combination, we did not find any significant interactions between group, word, and manipulation valence for CH rates (*b*

= -0.082, *SE* = 0.078, *z* = -1.051, *p* = 0.293) or v values (*b* = -0.146, *SE* = 0.133, *z* = -1.093, *p* = 0.274). However, patients had numerically higher CH rates and v values for all valence combinations. Notably, we did see a trend towards patients having higher CH rates and v values compared to controls in the stress manipulation-negative word valence condition (see figures 3.c and 3.d).

*Valence Impacts Confidence Ratings*

With respect to confidence ratings, we found a significant main effect of manipulation valence on confidence ratings. We found higher ratings in the stress condition compared to the non-stress condition for trials at 25%, 50%, and 75% intensity when indicating hearing a sound (25% trial: *b* = 0.16, *SE* = 0.07, *z* = 2.42, *p* = 0.016; 50% trial: *b* = 0.122, *SE* = 0.045, *z* = 2.744, *p* = 0.006; 75% trial: *b* = 0.087, *SE* = 0.030, *z* = 2.888, *p* = 0.004) and trials at 50% and 75% for all responses types (50% trial: *b* = 0.092, *SE* = 0.035, *z* = 2.610, *p* = 0.009; 75% trial: *b* = 0.084, *SE* = 0.030, *z* = 2.808, *p* = 0.005*).* The trend was the same for the trials with no word presented, although the results were not significant. For CH rates and v split by manipulation and word valence by manipulation type, please see figures S2-S3.

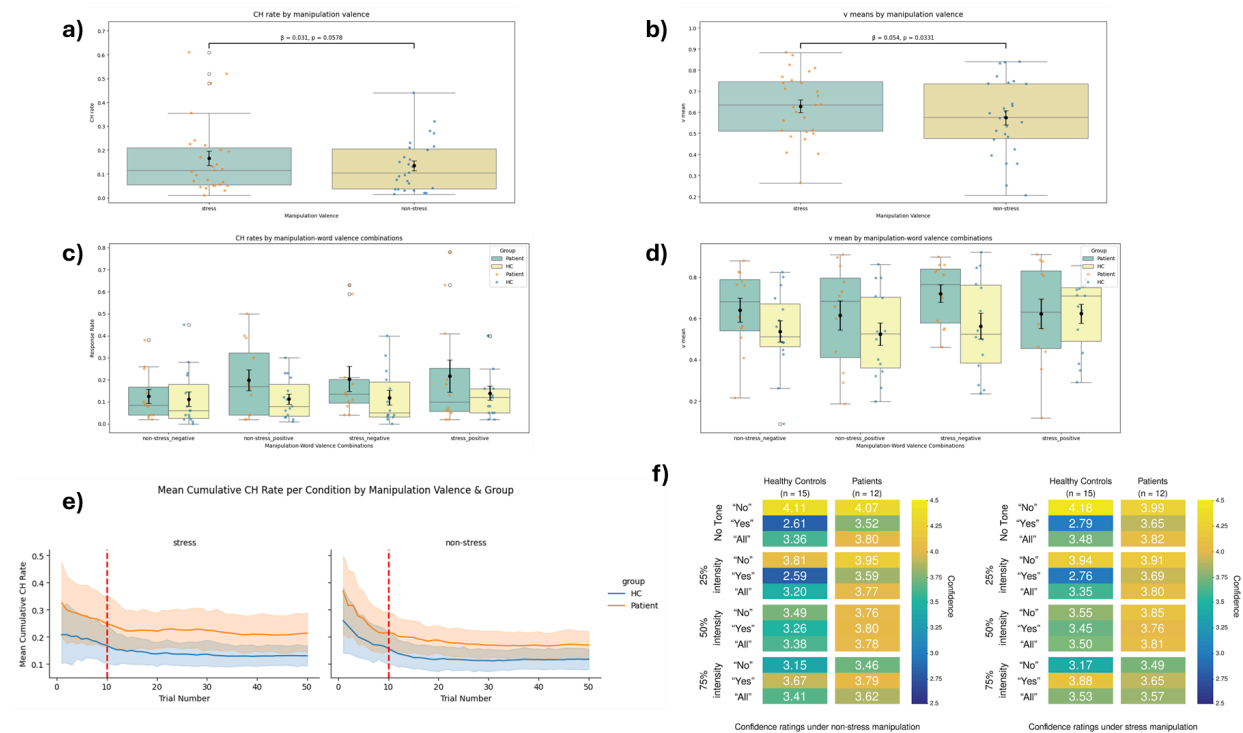

**Figure 3: Effect of a valenced manipulation and prior on CH rates and v values**

Panel (a) indicates CH rate by manipulation valence where boxplots depict group medians with 95 % confidence intervals (CIs) for the median, while black points with error bars indicate group means ± standard error of the mean (SEM). Individual points represent participant-level average CH rates. Panel (b) indicates v means by manipulation valence where boxplots show group medians with 95 % CIs for the median, and overlaid black points represent group means ± SEM. Panel (c) indicates CH rates by manipulation × word valence combinations boxplots display medians with 95% confidence intervals (CIs) for the median, while black points showing group means ± SEM across all combinations of manipulation (stress and non-stress) and word valence (negative and positive). Panel (d) indicates v means by manipulation × word valence combinations boxplots display medians with 95% confidence intervals (CIs) for the median, while black points showing group means ± SEM across all combinations of manipulation (stress and non-stress) and word valence (negative and positive).  Panel (e) shows the mean cumulative proportion of "yes" responses in no word trials plotted over trial number for healthy controls (HC) and patients for stress and non-stress manipulations. Shaded bands representing 95 % confidence intervals around the mean. The vertical red dashed line at trial 10 marks the point at which the manipulation occurred. Panel (f) shows confidence ratings by group, word presence, and intensity in stress and non-stress conditions. Heatmap showing mean confidence ratings (color-coded by magnitude) for "Yes," "No," and "All" responses across word-intensity levels (0 %, 25 %, 50 %, 75 %) and groups.

## *Effect of affective reaction on CH rates and v*

### *Mood change scores predict CH rates, with differential effects in patients and controls*

We found a main effect of normalized IMS rating change on CH rates ($b$ = 0.048, $SE$ = 0.052, $z$ = 2.557, $p$ = 0.011), a significant interaction between normalized IMS rating change and group as predictors for CH rates ($b$ = 0.032, $SE$ = 0.032, $z$ = -3.026, $p$ = 0.002), but no main effect of group ($b$ = 0.080, $SE$ = 0.052, $z$ = 1.553, $p$ = 0.121). See figure 4a-e for the full analysis and trends categorized by group.This demonstrates that patients showed a decrease in CH rates as overall change in total IMS score increased, and controls showed the opposite. Overall, we see that the effect of normalized IMS rating change and interaction between normalized IMS rating change and group on CH rates was driven primarily by the non-stress runs, potentially due to an emotional blunting response in patients. When looking at arousal change before and after the manipulation grouped across all run types per group, we saw that as arousal scores went up in the patient group, CH rates went down and vice versa in the control group. For more details on the analysis of IMS and arousal items, please see the supplementary materials.

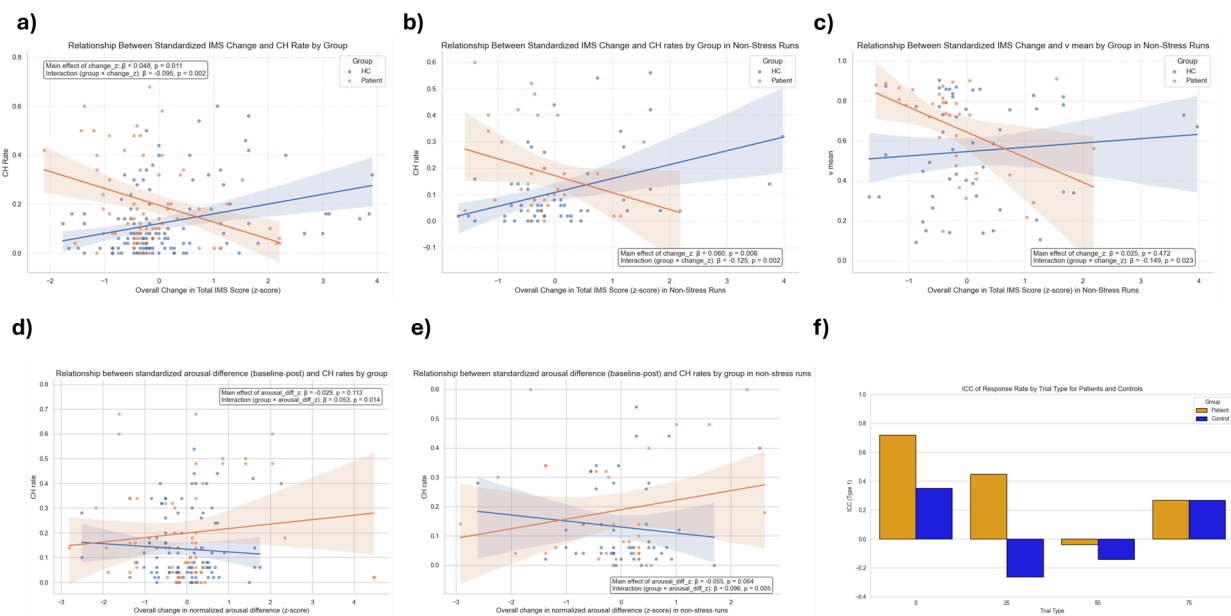

**Figure 4: Relationship between emotional ratings/arousal scores and CH rates/v values**
For panels a-e, scatterplots with regression lines illustrating the associations between standardized changes in total IMS/arousal scores (z-scored; *x*-axis) and CH rates/v values (*y*-axis) across healthy controls (HC; blue) and patients (orange). Each point represents an individual run for each participant, with shaded regions denoting 95% confidence intervals around the regression line. Positive *z*-scores indicate greater increases in IMS and arousal scores relative to the sample mean, whereas negative values reflect greater decreases. IMS means were calculated by taking the baseline IMS score (measured at the beginning of the session) and subtracting the IMS score from the end of each run (or run type). Arousal means were calculated by subtracting the values obtained from the VAS administered before the manipulation from the values obtained from the VAS administered after the manipulation. Panel (a) shows the relationship between CH rates and normalized IMS scores across all run types, panel (b) shows the relationship between CH rates and normalized IMS scores in the non-stress condition, panel (c) shows the relationship between v values and normalized IMS scores in the non-stress condition, panel (d) shows the relationship between CH rates and normalized arousal scores across all run types, and panel (e) shows the relationship between CH rates and normalized arousal scores in the non-stress condition. Panel (f) shows the intraclass correlation coefficients (ICC; Type 1) of response rates across sessions for each trial type (0%, 25%, 50%, 75%) in patients and healthy controls. Bars represent the stability of individual response rates within each group, with higher ICCs indicating greater within-subject consistency

### *Correlation between symptom scores with CH rates and v*

*LSHS scores were positively correlated with v values*

We aimed to replicate findings from Benrimoh et al. 2023 that showed there was a positive correlation between CH rates and v values with psychotic features characterized by LSHS scores with CH rates and v for session 1. We found, as expected, a significant positive correlation between CH rates and LSHS scores ($\rho$ = 0.42, *p* = .023; see figure S17); and a significant positive correlation between v values and LSHS scores ($\rho$ = 0.47, *p* = .014; see figure S18). See supplementary for an analysis of AHRS scores, which showed differing patterns.

### *Assessing intrarater reliability for CH rates and v within and between session*

*Patients compared to controls showed strong intrareliability between sessions for CH rates, but not for v values*

We found that CH rates for individual participants across session 1 and 2 had a relatively strong correlation of 0.66, which was found to be statistically significant (F = 4.907, *p* = 0.001). This effect was driven by patients who had a strong correlation of 0.83 in CH rates between sessions (F = 10.531, *p* = 0.003). Controls who had a weaker correlation of 0.34 in CH rates between sessions (F = 2.030, *p* = 0.143).

However, we found that v values for individual participants across sessions had a weak correlation of 0.18 and this was not statistically significant (F = 1.440, *p* = 0.232). These results indicate that patients had more consistent CH rates and prior weighting than controls between sessions; however, these results are limited by the relatively smaller number of patients who completed session 2 compared to controls.

We then examined the ICC for CH rates and nu between runs *within* a session, with the expectation that ICCs would be lower, given our hypothesis that differently valenced runs would produce different CH rates and prior weighting. Consistent with this hypothesis, we found relatively weak (but significant, indicating some effect of participant) correlations between runs for CH rates and v values in session 1 (CH rates session 1 - ICC: 0.40, F = 4.385, $p$ < 0.001; v values session 1 - ICC: 0.25, F = 2.627, $p$ = 0.003). Session 2 results, limited by a smaller number of participants, can be found in the supplementary material.

We next looked at the correlation between response rates in different trial types between sessions and found that the no-word trials had the strongest correlation value and was the only correlation that was significant (0% trial type ICC: 0.60, F = 3.961, $p$ = 0.003; see supplemental materials for other trial type correlations). We found that this effect was driven mostly by patients where there was a strong positive correlation in the no-word trials (significant) and 25 (non-significant) trials for the patients; these correlations were weaker and non-significant in the control group (0% trial type ICC for patients: 0.72, F = 6.093, $p$ = 0.016 - 0% trial type ICC for controls: 0.35, F = 2.078, $p$ = 0.135; 25% trial type ICC for patients: 0.44, F = 2.619, $p$ = 0.117 - 25% trial type ICC for controls: -0.26 F = 0.585, $p$ = 0.783; 50% trial type ICC for patients: -0.04, F = 0.928, $p$ = 0.528 - 50% trial type ICC for controls: -0.14, F = 0.752, $p$ = 0.661; 75% trial type ICC for patients: 0.27, F = 1.736, $p$ = 0.243 - 75% trial type ICC for controls: 0.27, F = 1.738, $p$ = 0.201). As such, patients demonstrate consistent conditioned hallucination rates, speaking to the reliability of the Aff-CH task to measure CH rates.

*Clinical Observations of Patients During the Task*

Patient experiences assessed during testing provide relevant information for understanding the effects of this task, our results, and to support future modifications of the task.

Two patients indicated experiencing reactivation of symptoms during the experiment, especially in the context of the negative stimuli; this did not include any new symptoms, only an increase in previously existing symptoms. One patient, who at baseline experienced thought broadcasting, indicated that, during a MIST stress run, this thought broadcasting worsened; this participant chose to end the experiment early as a result. Another patient who completed the first session of the experiment indicated hearing voices (yelling) when the white noise was present that distracted her from hearing the word that was a part of the experiment. This was worse after the negative than the positive AVAMs. The patient indicated having experienced similar symptoms regularly at home; for example when having a radiator on in her bathroom that has a similar noise she indicated hearing similar voices. Both of these patients were assessed by a psychiatrist after reporting these symptoms, and both indicated that their symptoms had returned to baseline within less than an hour of ending their session, and neither patient required hospitalization or emergent treatment. Both of these participants were removed from the study for their safety, and both of these incidents were reported as adverse events to the research ethics board. Another patient who regularly experienced voices noted that her symptoms interacted with the AVAM stimuli presented by commenting on them. She noted the negative stimuli made her think of negative things in the past that she felt the need to face, and the positive stimuli made her feel hopeful. This was not experienced as a worsening of symptoms and as such was not considered an adverse event. In summary, the affective stimuli qualitatively do seem to interact with positive symptoms; in particular, they seem to increase or engage existing symptoms, consistent with our hypothesis of affective states interacting with priors to worsen underlying symptoms. Given that similar events have not been reported, to our knowledge, with previous versions of the CH task, we suggest that these events were related to the uniquely affective nature of this task.

Further, patients, including those with reactivation of symptoms, often reported feeling sleepy after the experiment. This is consistent with our results showing a negative correlation between arousal scores and hallucination rates- where patients with lower arousal had higher CH rates and vice versa, indicating that some emotionally sensitive patients may react to affective stimuli with increased distance from their emotional experiences, experienced as sleepiness. In addition, some of the low CH rates in our patients may have been caused by distraction- e.g. those who experienced high arousal may have been overly attentive to the stimuli, to the point where it caused loss of focus on the task and reduced CH rates.

Indeed, when looking at the hallucination rates in patients who did have symptom reactivation, we found that they had very low conditioned hallucination rates across the whole session. As such, when patients are experiencing active symptoms in the context of the experiment, their attention may be focused more on their symptoms, distracting them from successfully completing the task. In addition, two patients with residual mania experienced agitation during the experiment and these two patients had low response rates to all trial types (not just low CH rates), indicating that agitation may have had an effect on attention in these patients, impacting responding across the experiment. These two patients were both outpatients at the time of testing and were able to provide informed consent; however their mania had not improved sufficiently with treatment to allow effective task engagement.

**Discussion**

We and others have argued that affective symptoms may be crucial in the development of psychosis, and predicted that affectively valenced tasks may accentuate phenomena such as conditioned hallucinations (D. Benrimoh et al., 2025); (Myin-Germeys & van Os, 2007). In the current pilot study we present the results of an affectively valenced version of the conditioned hallucinations task, combining both valenced priors (auditory stimuli) and stress vs. non-stress manipulations. The primary objective of this pilot study was to validate that adding affective stimuli in this way to the CH task does not invalidate it. Our results showed a replication of the increase in conditioned hallucination rates (at the first session) and prior weighting (across all runs) in patients compared to controls, and this finding was numerically consistent across all run types. In addition, the shape of the cumulative CH rate graph in figure 1.d is essentially identical to that seen in (D. Benrimoh et al., 2024; Kafadar et al., 2022), indicating that adding the affective manipulation does not cause an accelerated decrease in prior strength, which was a significant concern when we decided to add a manipulation *during* the CH task. We also showed that the v and CH rates correlated with each other and with symptoms, consistent with previous work (D. Benrimoh et al., 2024; Kafadar et al., 2022). Finally, we established significant test-retest reliability for the task, which was strongest for patients.

Our second objective was to demonstrate that affect could indeed influence prior weighting in this task, with the explicit hypothesis that the priors underlying clinical hallucinations are often valenced in nature (often negatively valenced), and that stressful experiences which engender affective states should be able to dynamically alter the weighting of these priors. Given the literature demonstrating that stress tends to increase positive symptoms (Hartley et al., 2013)(Lysaker & Salyers, 2007), we expected that stress manipulations and resulting negative affective states would have stronger effects on prior weighting. Importantly, we demonstrated that stress runs, compared to non-stress runs, did in fact have higher prior weighting; for CH rates this was non-significant at p = 0.058. While not significant in this underpowered study, we also did observe that, across all runs and sessions, the numerically highest prior weighting and CH rates were observed in patients in the negative word-stress manipulation conditions. In addition, the largest between-group differences in cumulative CH rates were seen in stress conditions. The finding of relatively lower ICCs for CH rates and v estimates within a session also supports the likely differences between differently valenced runs. One key objective of future studies, currently in planning, will be to confirm these differences.

Our findings on the IMS and arousal score pre-post manipulation suggest different processing and reporting of emotional experience in patients and controls; table S21 also suggests relative blunted arousal change in patients vs controls in the stress condition. This is consistent with the literature, which shows blunted reported emotional reports in patients with psychosis after affective stimulus presentation (Aminoff et al., 2011); (Liu et al., 2022); Kring and Moran 2008), which is hypothesized to be related to different experiences of and reporting of emotion in the psychosis spectrum (Peterman et al. 2015; Gooding and Tallent 2002; (Aminoff et al., 2011). Interestingly, in both analyses the group by score change interaction was apparent in the non-stress runs, but not in the stress runs. As such, it may be that stress runs make patients and controls seem "more alike", aligning with our hypotheses about the specific importance of negative affect. In addition, many patients noted to us that they felt tired after the experiment (see qualitative section). This may reflect the taxing nature of the experiment on patients, who may have low energy due to depressive or negative symptoms. However, an intriguing possibility, arising from our clinical experience and the literature (Denollet & De Vries, 2006) is that fatigue may be a sign of dissociation or a reaction to negative affective experience. As such, the patients with lower arousal post-manipulation and associated higher prior weighting may have had a more significant affective reaction to the stimuli, but may not be able to report it as such due to differences in how emotions are experienced and reported in patients compared to controls. Physiological monitoring during the experiment, and larger sample sizes which would allow for the identification of emotion processing subgroups within the patient population, would help to clarify if these hypotheses have merit.

Based on qualitative reports, and with a view to producing a more tolerable version of this task, we decided to alter the task for future studies. First, in order to reduce the risk of overly salient affective stimuli stimuli, we decided to remove the negative AVAMs from future studies. This was motivated both by patients and controls reporting being more emotionally activated by the AVAMs, and by results (see supplementary) suggesting that the MIST produced higher CH rates than the AVAMs, potentially because it has less of a disruptive impact on attention (see analysis of attention IMS item in supplementary materials, table S19) . We also chose to limit the word stimuli to clearly valenced but less aversive words (e.g. keeping "bad" and "stress" and removing words such as "die"). In addition, we argue that it is crucial to provide patients completing these tasks with the opportunity to reach out for support if they experience any negative impacts, and to have procedures in place for offering resources.

This study has a number of significant limitations. First and foremost is its small sample size, which likely led to a failure to detect a number of effects while increasing the uncertainty of the effects observed. Future studies will require larger samples in order to confirm and extend our findings. Another significant limitation is the lack of a non-affectively valenced comparator condition. We chose not to employ the classic CH task as a comparator as we were concerned that the inclusion of the manipulation might have changed task dynamics, making the comparison inappropriate. We are currently in the process of testing and designing a neutral condition and we expect to publish those results separately once more data has been collected. Another limitation is our heterogenous patient sample, which included patients at varying stages of psychosis risk and early psychosis, and only required auditory hallucination symptoms in the last three months (which may have diluted effects, given the sensitivity of the CH task to hallucination status(Kafadar et al., 2022)). Given the CH task is sensitive to illness state (Kafadar et al., 2022), this may have reduced the signal in our already small patient group; however, given that patients in our clinics are treated very quickly for hallucinations and these often respond rapidly to antipsychotic medication, this sampling approach was a practical necessity for this pilot sample, and it is encouraging to note the initial suggestions of an effect of affect in the patient group despite its small size and heterogeneity. Finally, the lack of physiological or neurobiological data limits the interpretation of affect and arousal change data; these are planned to be added in the next phase of this work.

Overall, the initial results support the hypotheses that prior weighting will be increased in negative affective contexts. Patients demonstrated higher prior weighting compared to healthy controls, and negative emotional manipulations contributed to overweighting of priors. The findings from this pilot study support the feasibility of developing a computational task capable of capturing how emotional context interacts with belief formation and perceptual inference. These tools provide mechanistically-informed markers of psychosis severity and may serve important purposes for early detection and personalizing interventions for individuals at risk for psychosis. Finally, targeting individuals whose hallucination risk is exacerbated by stress and focusing interventions on emotional regulation and belief flexibility may offer a novel step forward for the field of early psychosis research.

**Supplementary Materials**

Table S1: Mean symptom scores and medication for first and second session split by patients with an FEP and CHR-P diagnosis

|  | **Patients with FEP diagnosis** | **Patients with CHR-P diagnosis** |
|---|---|---|
| **AHRS scores** | Mean: 23.400<br>Std: 7.701 | Mean: 13.714<br>Std: 6.993 |
| **LSHS scores** | Mean: 22.400<br>Std:13.465 | Mean: 28.000<br>Std: 7.118 |
| **PHQ-9 scores** | Mean: 7.200<br>Std: 6.834 | Mean: 12.857<br>Std: 4.337 |
| **GAD-7 scores** | Mean: 9.400<br>Std: 5.595 | Mean: 12.000<br>Std: 5.831 |
| **Medication** | Were taking antipsychotics for the first session: 100% (5/5)<br><br>Were taking antipsychotics for the second session: 100% (2/2) | Were taking antipsychotics for the first session: 43% (3/7)<br><br>Were taking antipsychotics for the second session: 50% (3/6) |
|  | **Session 1** | **Session 2** |
| **002 - CHR-P** | Risperidone, Sertraline, Dexedrine, Xanax | Risperidone, Lurasidone, Propranolol, Sertraline, Dexedrine, Xanax |
| **006 - CHR-P** | No medication | No medication |
| **007 - CHR-P** | Risperdine, Olanzapine, Adderal | Adderal |
| **009 - CHR-P** | No medication | No medication |
| **012 - CHR-P** | Olanzapine, Lamotrigine | Olanzapine, Risperidone |
| **015 - FEP** | Lurasidone, Escitalopram, Gabapentin, Olanzapine | No session 2 |
| **020 - CHR-P** | Information not present | No session 2 |
| **023 - FEP** | Latuda, Quetiapine, Fluoxetine | No session 2 |
| **027 - CHR-P** | Bupropion | Bupropion, Quetiapine |
| **029 - FEP** | Citalopram, Abilify | Abilify |
| **030 - FEP** | Abilify | No session 2 |

| | | | |
|---|---|---|---|
| 031 - FEP | Aripiprazole maintena | Aripiprazole maintena | |

## Replication of original CH task - supplementary section

We found that the correlation between CH rates and *v* values was driven by the patient group who showed a stronger positive correlation (ρ = 0.545, *p* = 0.016) where the control group showed a weaker, non-significant, positive correlation (ρ = 0.175, *p* = 0.403). Looking at the range of CH rates and v values for patients and controls grouped across sessions and runs, we see that patients have higher minimum and maximum CH rates and v values compared to controls (patient min CH rate: 0.047, patient max CH rate: 0.525, HC min CH rate: 0.015, HC max CH rate: 0.318; patient min v value: 0.363, patient max v value: 0.817, HC min v value: 0.310, HC max v value: 0.757). Furthermore, as in (A. R. Powers et al., 2017), we found that as the audibility increased, the response rates of the patient group approached the values of the control groups, as shown in Figure 2.c. We found a significant main effect of group on response rates in the 25% (*b* = -0.109, *SE* = 0.051, *z* = -2.140, *p* = 0.032) and 50% audibility conditions (*b* = -0.042, *SE* = 0.002, *z* = -25.286, *p* < 0.001) where patients showed higher response rates compared to controls, but not in the 75% audibility condition (*b* = 0.019, *SE* = 0.048, *z* = 0.392, *p* = 0.695) where controls showed numerically higher response rates, but no significant group differences were found.

*Confidence ratings:* We looked at the effect of different audibility trial types between groups for confidence rates. We used linear mixed effect models with group as a fixed effect and random intercepts for participants to account for within-subject variability. We found no significant main effect of the group on confidence ratings. We found a numerical trend of patients rating being higher than HC for trials where there was an indication of hearing a sound on no-word trials, 25% intensity trials, and 50% intensity trials (See fig 2.e). The model included a random intercept variance of around 1.1 across trial types, indicating a substantial variability in overall rating levels in participants.

### CH rates by group

Table S2: CH rates overall by group

| Patient CH rates statistics | Patient n | HC CH rates statistics | HC n |
|---|---|---|---|
| Mean: 0.187<br>Std: 0.161 | 12 | Mean: 0.120<br>Std: 0.088 | 15 |

### v values by group

Table S3: v values overall by group

| Patient v value statistics | Patient n | HC v value statistics | HC n |
|---|---|---|---|
| Mean: 0.655<br>Std: 0.159 | 12 | Mean: 0.560<br>Std: 0.141 | 15 |

**Response rates by trial type and group**

Table S4: Response rates by trial type and group

| Trial type | Patient response rates | HC response rates |
|---|---|---|
| 25 | Mean: 0.365<br>Std: 0.170 | Mean: 0.256<br>Std: 0.010 |
| 50 | Mean: 0.5610<br>Std: 0.2260 | Mean: 0.520<br>Std: 0.170 |
| 75 | Mean: 0.797<br>Std: 0.148 | Mean: 0.816<br>Std: 0.120 |

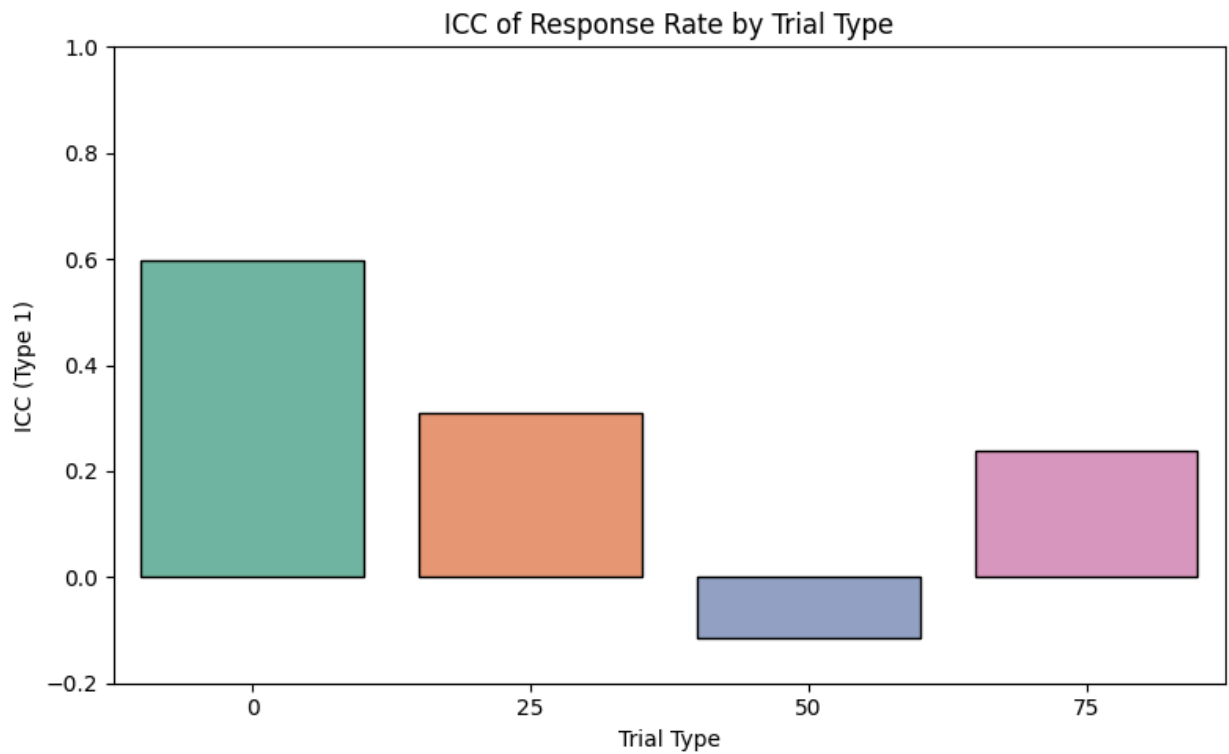

Figure S1: ICC by trial type

*ICC for response rate by trial type per group*

We looked at the ICC coefficient for different trial types per group testing the consistency in response for specific trial types across the two sessions. We found that in the patient group, for no word trials, there was a strong positive correlation of 0.60 which was statistically significant (F = 3.961, *p* = 0.003), for the 25 trial type, there was a moderate positive correlation of 0.31 (F = 1.894, *p* = 0.101), for the 50 trial type, there was a weak negative correlation of -0.11 (F = 0.796, *p* = 0.674), and for the 75 trial type, there was a weak positive correlation of 0.24 (F = 1.629, *p* = 0.164).

For the HC group, for no word trials, there was a moderate positive correlation of 0.35 (F = 2.078, *p* = 0.135), for the 25 trial type, there was a moderate negative correlation of -0.26 (F = 0.585, *p* = 0.783), for the 50 trial type, there was a weak negative correlation of -0.14 (F = 0.752, *p* = 0.661), and for the 75 trial type, there was a moderate positive correlation of 0.27 (F = 1.738, *p* = 0.201). Patients had a numerically stronger correlation value of 0.27 for v values between sessions (F = 1.743, *p* = 0.242) compared to controls who had a numerically weaker correlation value of -0.004 for v values between sessions (F = -0.993, *p* = 0.500). However, correlation values overall, and by group were not statistically significant for v values between sessions.

Table S5: CH rates by group and manipulation valence

|  | Patient CH rates statistics | HC CH rates statistics |
| --- | --- | --- |
| Stress | Mean: 0.220<br>Std: 0.233 | Mean: 0.130<br>Std: 0.141 |
| non-stress | Mean: 0.178<br>Std: 0.162 | Mean: 0.118<br>Std: 0.144 |

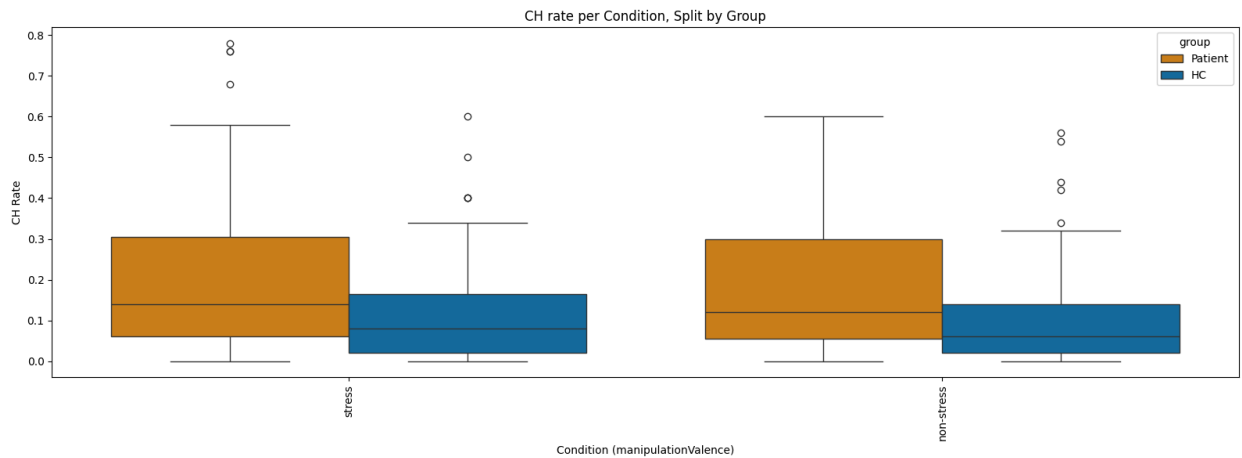

Figure S2: CH rates by group and manipulation valence

### *Effect of manipulation valence by group*

We looked at the effect of the stress vs. non-stress manipulations between groups on CH rates and prior weighting grouped across sessions. We found that CH rates and v values were numerically higher in the patient compared to the control group in all run types (see table S9). When looking at CH rates between groups, we did not find a significant interaction between group and manipulation valence ($b$ = 0.031, $SE$ = 0.033, $z$ = 0.959, $p$ = 0.338). When looking at cumulative CH rates post manipulation, we found that patients had numerically higher cumulative CH rates in both stress and non-stress conditions (see figure S4).

Table S6: v values by group and manipulation valence

|  | Patient v statistics | HC v statistics |
|---|---|---|
| Stress | Mean: 0.693<br>Std: 0.231 | Mean: 0.606<br>Std: 0.2407 |
| Non-stress | Mean: 0.648<br>Std: 0.232 | Mean: 0.553<br>Std: 0.252 |

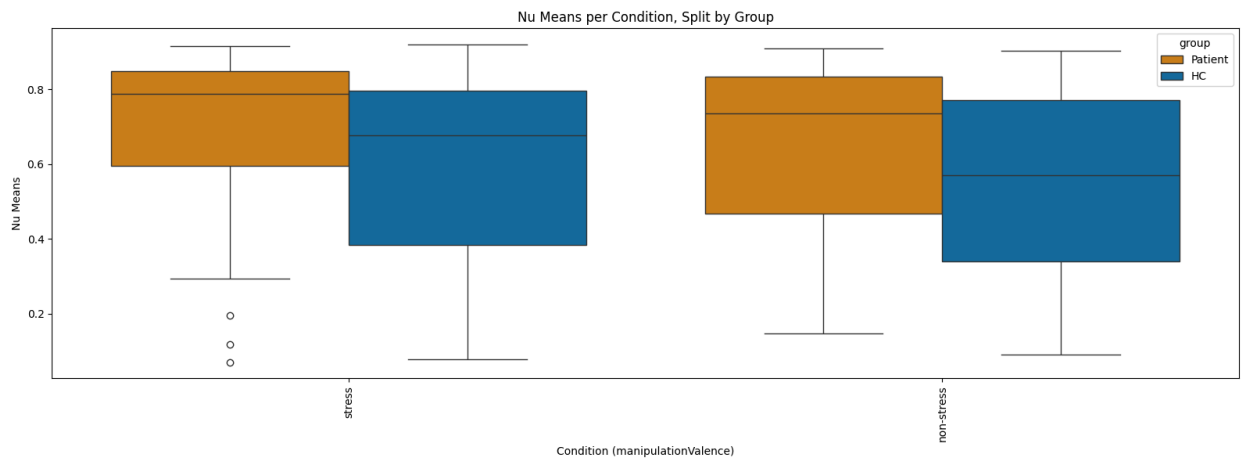

Figure S3: v values by group and manipulation valence

Table S7: CH rates by group and word valence

|  | Patient CH rates statistics | HC CH rates statistics |
| --- | --- | --- |
| Negative | Mean: 0.185<br>Std: 0.189 | Mean: 0.117<br>Std: 0.1431 |
| Positive | Mean: 0.214<br>Std: 0.162 | Mean: 0.131<br>Std: 0.142 |

Table S8: v values by group and word valence

|  | Patient v statistics | HC v statistics |
| --- | --- | --- |
| Negative | Mean: 0.701<br>Std: 0.201 | Mean: 0.562<br>Std: 0.258 |
| Positive | Mean: 0.641<br>Std: 0.257 | Mean: 0.594<br>Std: 0.237 |

Table S9: CH rates by group and manipulation-word valence combinations

|  | Patient CH rates statistics | Patient n runs | HC CH rates statistics | HC n runs |
| --- | --- | --- | --- | --- |
| Stress-Negative | Mean: 0.224<br>Std: 0.227 | 19 | Mean: 0.121<br>Std: 0.1392 | 23 |
| Stress-Positive | Mean: 0.216<br>Std: 0.244 | 19 | Mean: 0.139<br>Std: 0.145 | 25 |
| Non-stress-Negative | Mean: 0.144<br>Std: 0.132 | 18 | Mean: 0.113<br>Std: 0.149 | 25 |
| Non-stress-Positive | Mean: 0.212<br>Std: 0.185 | 18 | Mean: 0.122<br>Std: 0.142 | 25 |

Table S10: v values by group and manipulation-word valence combinations

|  | Patient CH rates statistics | Patient n runs | HC CH rates statistics | HC n runs |
| --- | --- | --- | --- | --- |
| Stress-Negative | Mean: 0.733<br>Std: 0.195 | 18 | Mean: 0.575<br>Std: 0.270 | 23 |
| Stress-Positive | Mean: 0.653<br>Std: 0.261 | 18 | Mean: 0.634<br>Std: 0.213 | 25 |

| | | | | | |
|---|---|---|---|---|---|
| Non-stress-Negative | Mean: 0.667 Std: 0.207 | 17 | Mean: 0.551 Std: 0.252 | 25 | |
| Non-stress-Positive | Mean: 0.629 Std: 0.259 | 18 | Mean: 0.555 Std: 0.257 | 25 | |

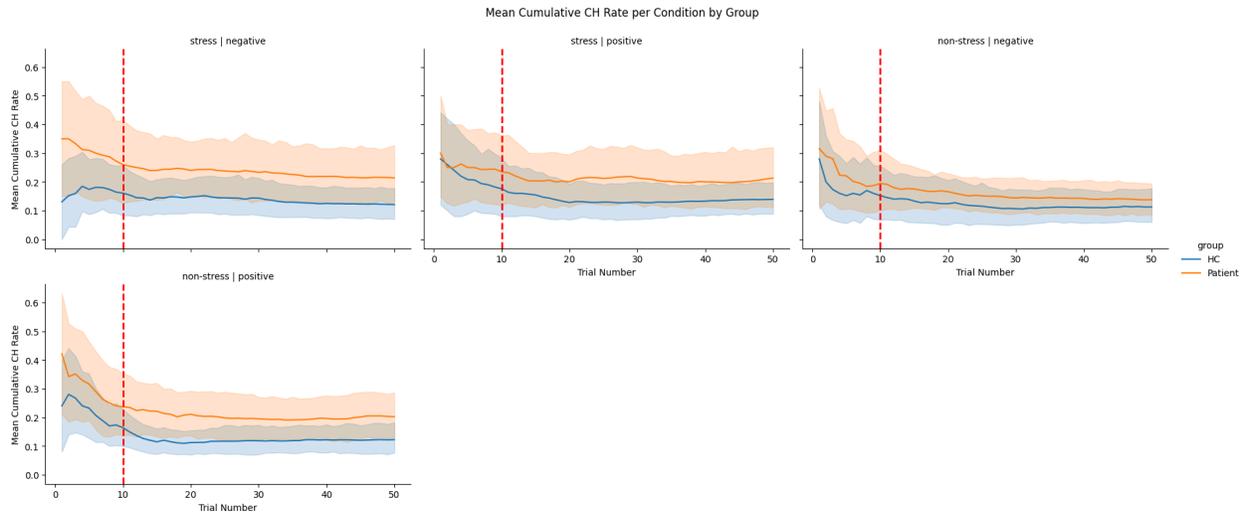

Figure S4: Cumulative CH rates by group and manipulation-word valence combinations

### CH rates and v by different manipulation types

We looked at the CH rates and v values across the different manipulation types to see if there were differences in behavioral and computational parameters between manipulation types. We first used Mann Whitney U tests to compare CH rates and v values from different manipulation types. We found a significant difference in CH rates and v values for the stress MIST versus the non-stress AVAM manipulations (CH rates: $U$ = 1269.5, $p$ = 0.011; v values: $U$ = 1249.0, $p$ = 0.018). We did not see a significant difference when looking at differences between the stress MIST and stress AVAM for CH rates nor v values, however, both were trending towards statistical significance (CH rates: $U$ = 1145.5, $p$ = 0.055; $U$ = 1134.0, $p$ = 0.070). Looking within session types, we used Wilcoxon Signed Ranked Tests. Looking at non-stress versus stress MIST, we did not see a significant difference in CH rates nor v values (CH rates: $U$ = 36.0, $p$ = 0.187; v values: $U$ = 40.0, $p$ = 0.277). Similarly, looking at the difference between non-stress and stress AVAM, we did not see a significant difference in CH rates nor v values (CH rates: $U$ = 69.0, $p$ = 0.495; v values: $U$ = 59.0, $p$ = 0.265).

Table S11: CH rates by manipulation type

| Manipulation Valence | Manipulation Type | Mean CH rates | Std CH rates | N runs |
|---|---|---|---|---|
| Stress | AVAM | 0.1400 | 0.1700 | 44 |
| | MIST | 0.2014 | 0.2085 | 42 |
| Non-stress | AVAM | 0.1217 | 0.1485 | 46 |
| | MIST | 0.1675 | 0.1586 | 40 |

Table S12: v values by manipulation type

| Manipulation Valence | Manipulation Type | Mean v values | Std v values | N |
|---|---|---|---|---|
| Stress | AVAM | 0.6029 | 0.2375 | 44 |
| | MIST | 0.6872 | 0.2359 | 42 |
| Non-stress | AVAM | 0.5634 | 0.2480 | 46 |
| | MIST | 0.6264 | 0.2446 | 40 |

***CH rates and v values by manipulation-word valence combinations for each manipulation type***

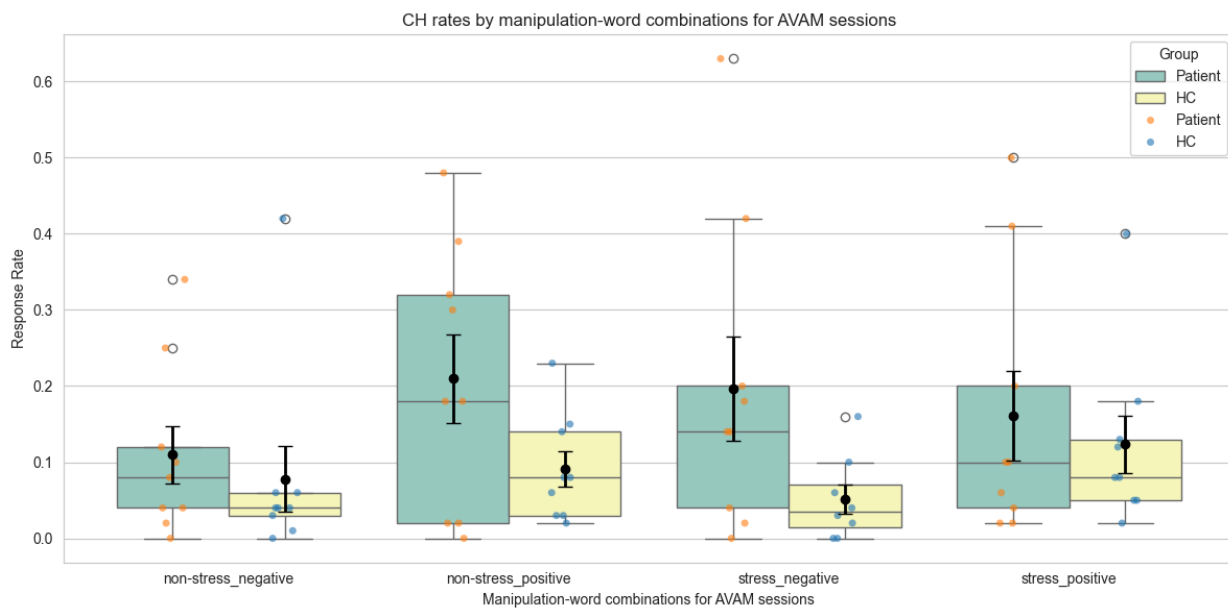

Figure S5: CH rates by manipulation-word valence combinations for AVAM sessions

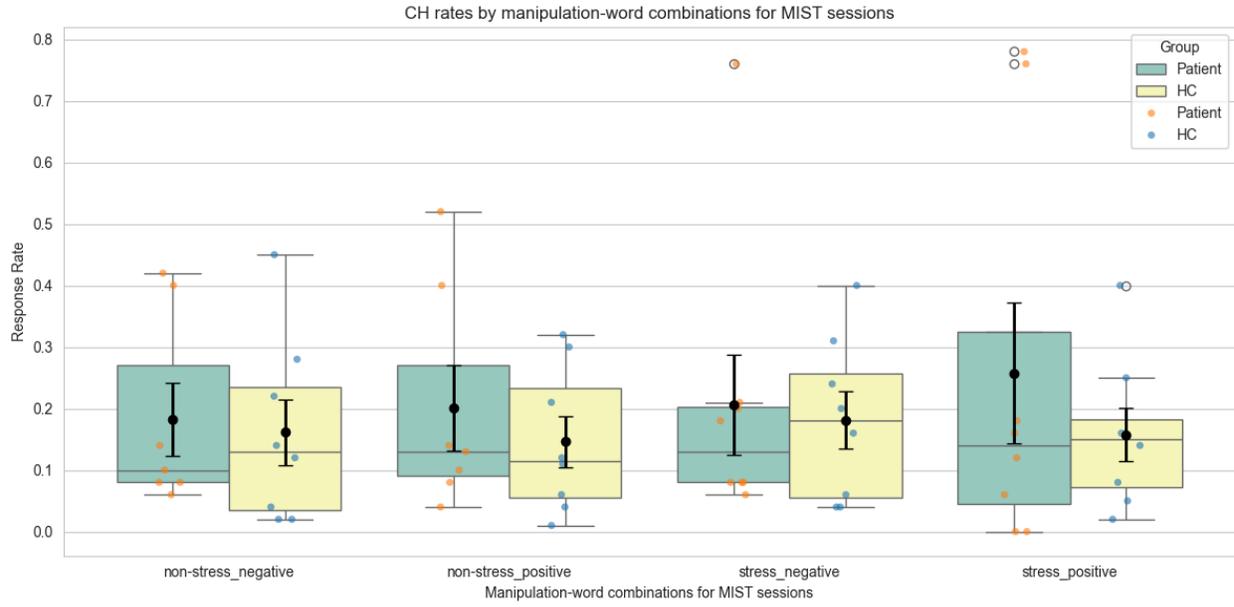

Figure S6: CH rates by manipulation-word valence combinations for MIST sessions

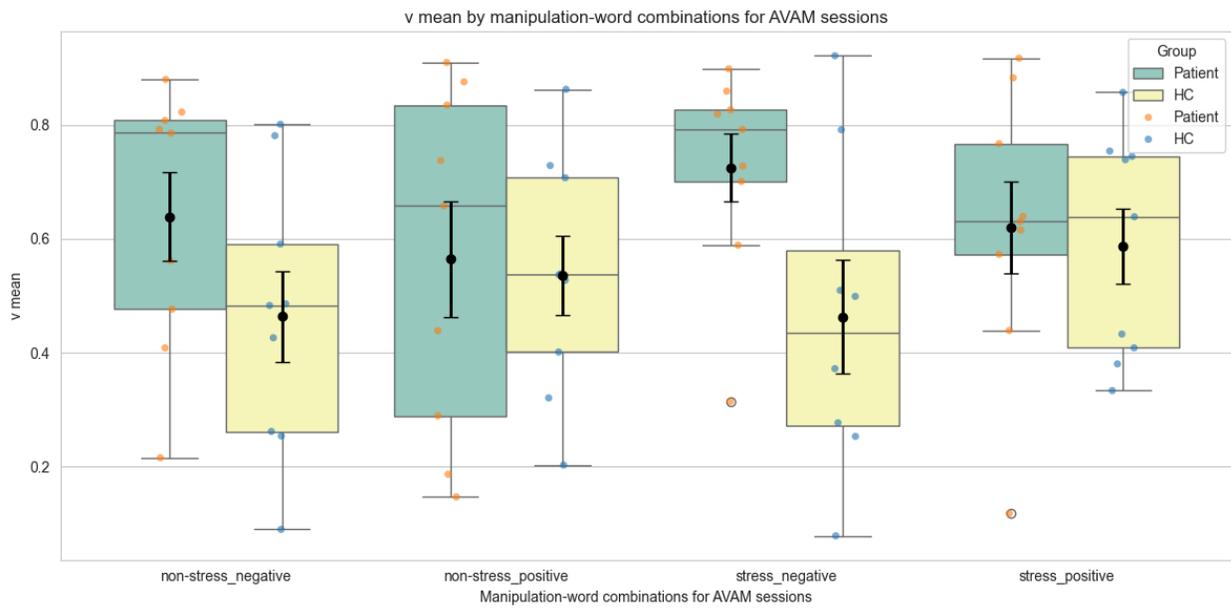

Figure S7: v values by manipulation-word valence combinations for AVAM sessions

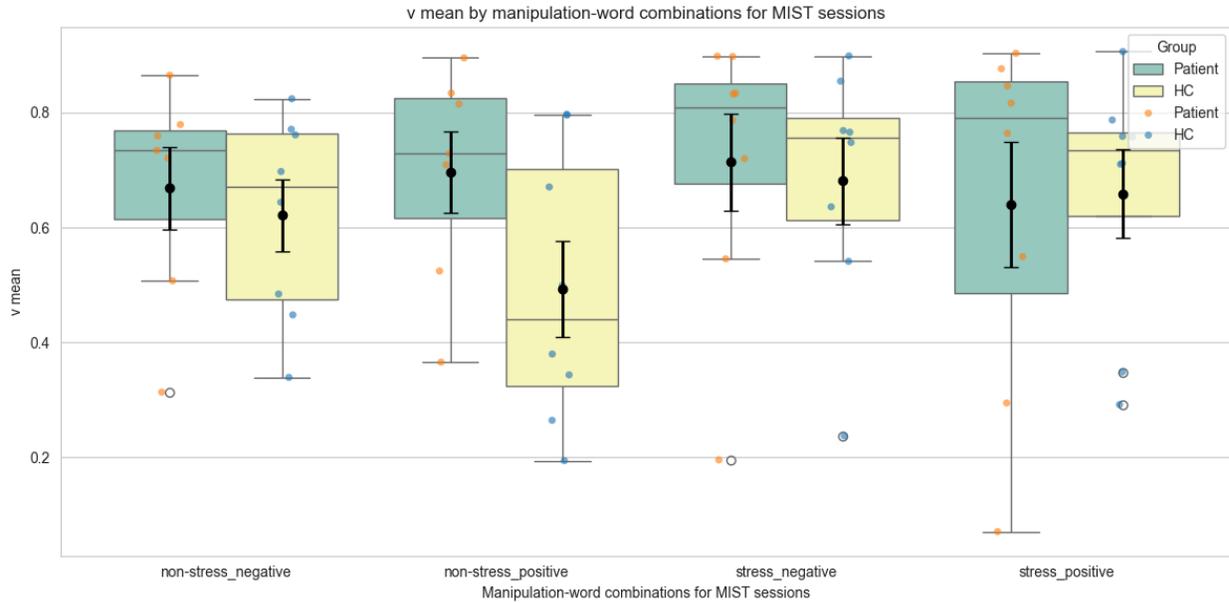

Figure S8: v values by manipulation-word valence combinations for MIST sessions

***Association between word (prior) valence with CH rates and v values by group***

We looked at the effect of negatively or positively valenced priors (words) on CH rates and v between groups. We found that for runs with words of positive valence in patients had numerically higher CH rates compared to controls, and runs with words of negative valence in patients had numerically higher v values compared to controls (mean CH rates for word conditions in patients - negative words: 0.209, positive words: 0.165; mean CH rates for word conditions in controls - negative words: 0.113, positive word: 0.126; mean v values for word conditions in patients - negative words: 0.680, positive words: 0.629; mean v values for word conditions in controls - negative words: 0.543, positive words: 0.574) ). Overall we did not see an effect of word valence on CH rates ($b$ = 0.013, $SE$ = 0.031, $z$ = 0.409, $p$ = 0.683) nor v values ($b$ = 0.031, $SE$ = 0.045, $z$ = 0.685, $p$ = 0.493) and we also did not see a significant interaction between word valence and group for CH rates ($b$ = 0.031, $SE$ = 0.046, $z$ = 0.683, $p$ = 0.495) nor for v values ($b$ = -0.082, $SE$ = 0.068, $z$ = -1.207, $p$ = 0.227).

We did not find a main effect of word valence on CH rates nor v values as well as not significant interactions by group. Please see tables S14 and S15 for the CH rates and v values by word (prior) valence and group.

Table S13: Mean CH rates by word (prior) valence in patients versus controls

| *CH rates* | **Patient** | **Control** |
|---|---|---|
| **Negative** | 0.209 | 0.113 |
| **Positive** | 0.165 | 0.126 |

Table S14: Mean v values by word (prior) valence in patients versus controls

| v values | Patient | Control |
|---|---|---|
| **Negative** | 0.680 | 0.543 |
| **Positive** | 0.629 | 0.574 |

***Nu means by block***

We ran the HGF model by the three different blocks in the experimental paradigm to examine the change in prior weighting in blocks before and after the manipulation. Examining group differences across the different blocks, we found that patients had significantly higher nu means in block 1, but this difference was not significant for blocks 2 and 3 while patients did still show numerically higher nu values for these blocks (Patient nu mean block 1: 0.688, control nu mean block 1: 0.610; $U = 5199.0$, $p = 0.001$ - patient nu mean block 2: 0.670, control nu mean block 2: 0.650; $U = 4693.0$, $p = 0.076$ - patient nu mean block 3: 0.682, control nu mean block 3: 0.666; $U = 4668.0$, $p = 0.088$). Looking at group differences in each block by manipulation valence, we found that patients had significantly higher nu values in block 1 for the non-stress condition, and this difference was trending towards statistical significance for the stress condition in block 1 (Patient nu mean block 1 stress condition: 0.686, control nu mean block 1 stress condition: 0.622; $U = 1246.0$, $p = 0.055$ - patient nu mean block 1 non-stress condition: 0.689, control nu mean block 1 non-stress condition: 0.598; $U = 1349.0$, $p = 0.011$). As for the other 2 blocks in different manipulation valence conditions, we did not see any significant group differences, however, patients showed numerically higher nu values in all blocks except for non-stress negative condition in block 3 where controls showed higher nu values (Patient nu mean block 2 stress condition: 0.681, control nu mean block 2 stress condition: 0.658; $U = 1211.0$, $p = 0.101$ - patient nu mean block 2 non-stress condition: 0.659, control nu mean block 2 non-stress condition: 0.641; $U = 1136.0$, $p = 0.385$ - patient nu mean block 3 stress condition: 0.677, control nu mean block 3 stress condition: 0.668; $U = 1127.0$, $p = 0.336$ - patient nu mean block 3 non-stress condition: 0.688, control nu mean block 3 non-stress condition: 0.665; $U = 1212.0$, $p = 0.141$).

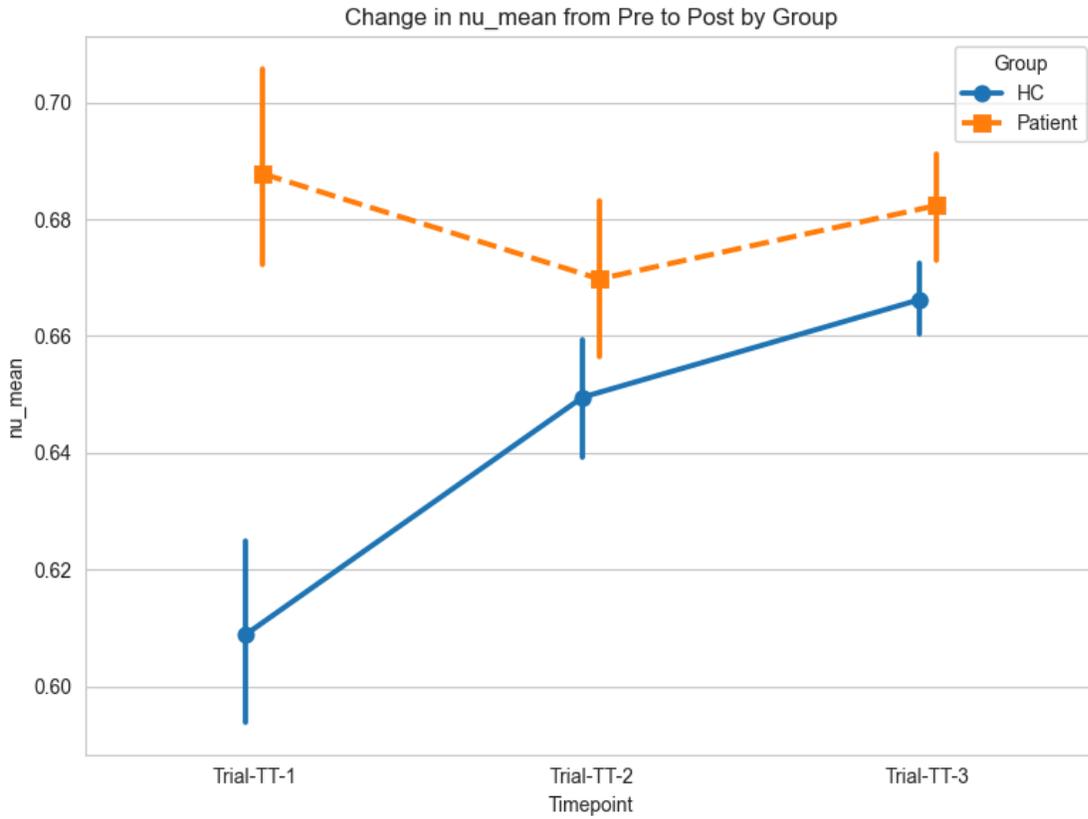

Figure S9: v mean by group and block

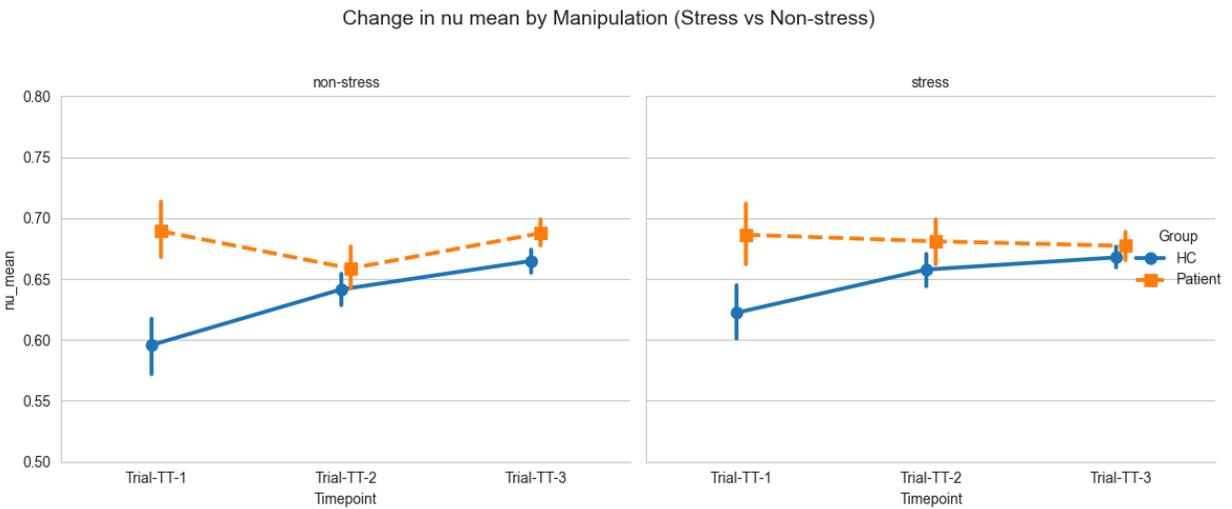

Figure S10: v mean by group and block for stress and non-stress manipulation valence

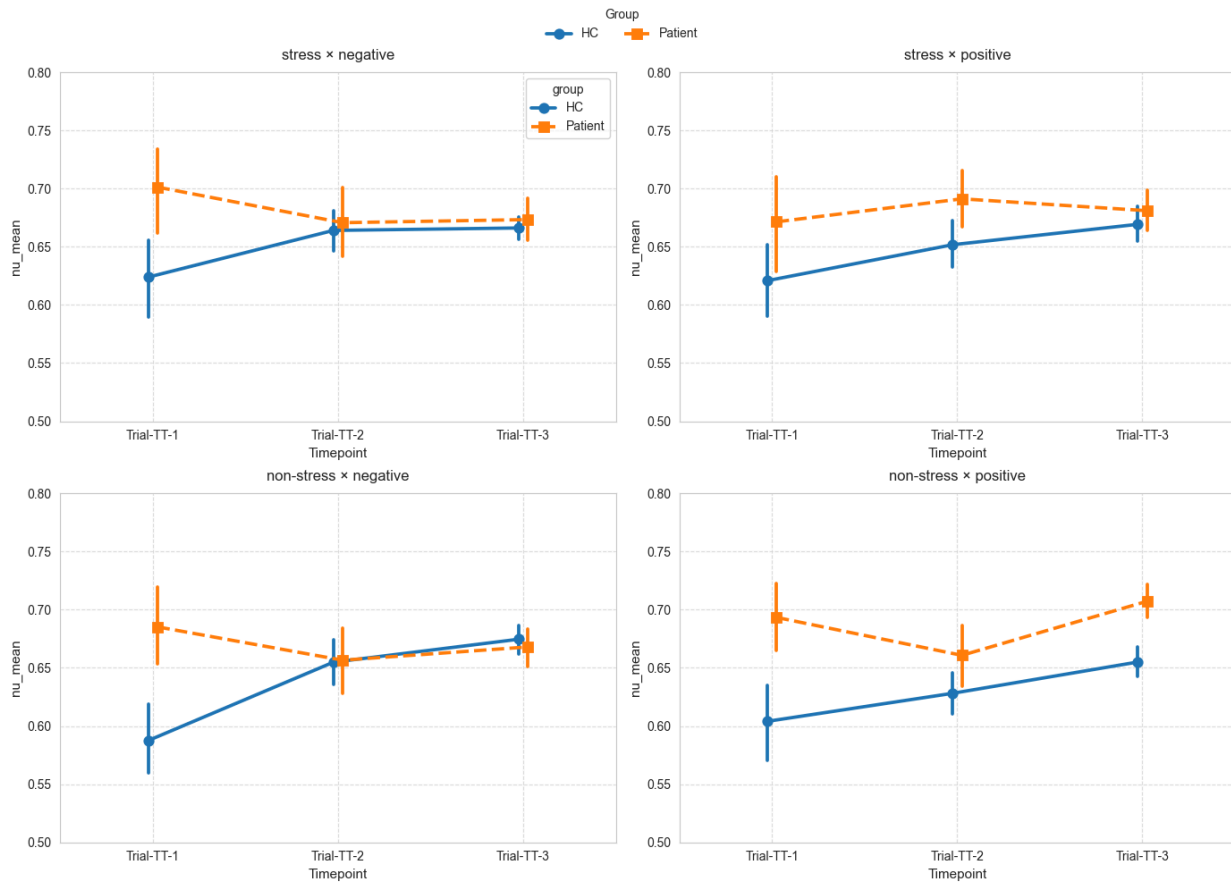

Figure S11: v mean by group and block for manipulation word valence combinations

## *Nu means by pre versus post manipulation*

We ran the HGF model by trials that were before and after the manipulation to examine the change in prior weighting due to the manipulation. Overall, we found that patients compared to controls had significantly higher nu values before, but not after the manipulation (Patient nu mean pre manipulation: 0.684, control nu mean pre manipulation: 0.609; $U$ = 6940.0, $p < 0.001$ - patient nu mean post manipulation: 0.667, control nu mean post manipulation: 0.643; $U$ = 6295.0, $p$ = 0.051). When looking at how the different manipulation valence conditions affect nu before and after the manipulation, we compared group differences for both pre and post manipulation conditions. We found significant group differences in the non-stress condition prior to the manipulation where patients had higher nu values compared to the control group, but not for the stress condition (Patient nu mean pre manipulation in stress condition: 0.686, control nu mean pre manipulation in stress condition: 0.622; $U$ = 1241.0, $p$ = 0.060 - patient nu mean pre manipulation in non-stress condition: 0.690, control nu mean pre manipulation in non-stress condition: 0.596; $U$ = 1348.0, $p$ = 0.011), however, we did not find significant group differences in either of the valence conditions after the manipulation (Patient nu mean post manipulation in stress condition: 0.671, control nu mean post manipulation in stress condition: 0.661; $U$ =

1149.0, *p* = 0.255 - patient nu mean post manipulation in non-stress condition: 0.658, control nu mean post manipulation in non-stress condition: 0.625; *U* = 1203.0, *p* = 0.0162).

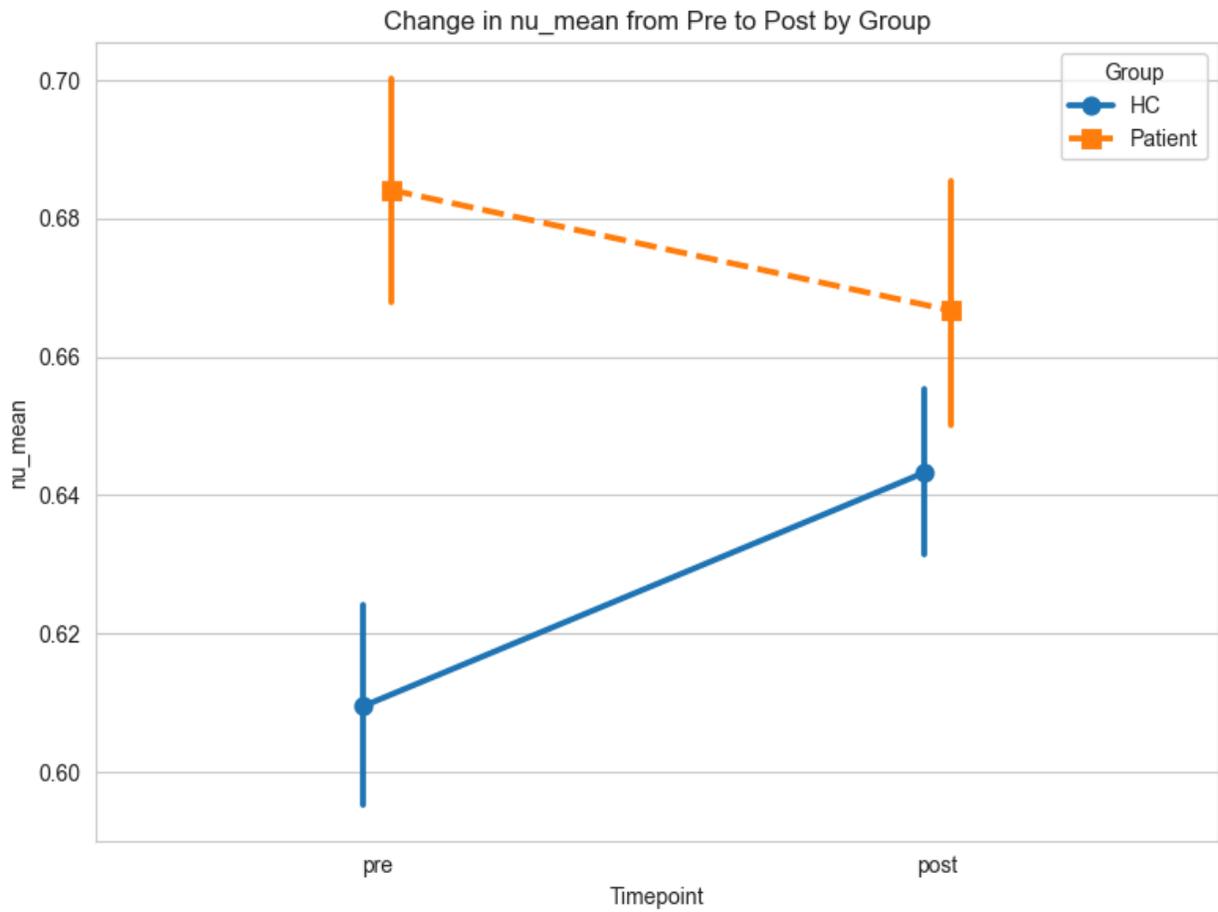

Figure S12: Change in v from pre to post manipulation by group

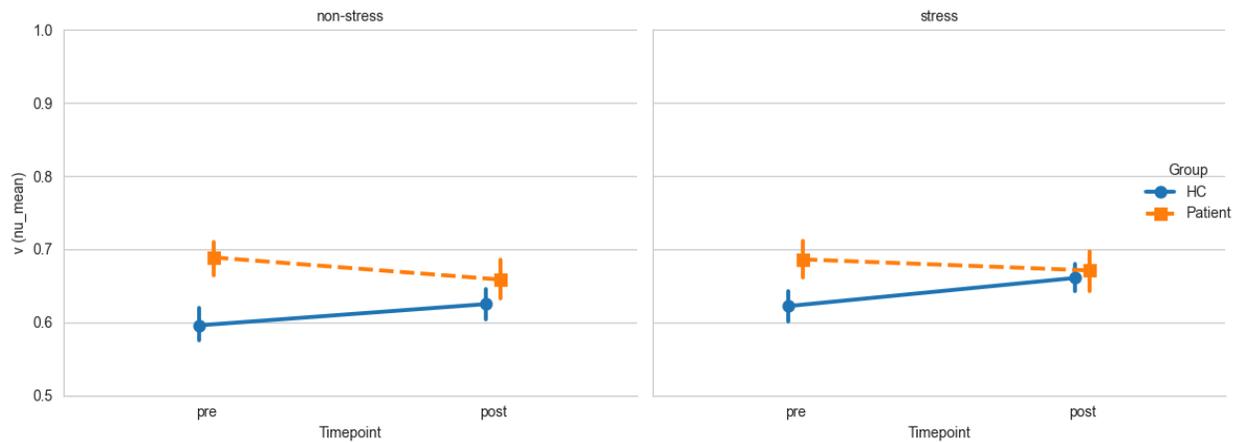

Figure S13: Change in v mean from pre to post manipulation by group and manipulation valence

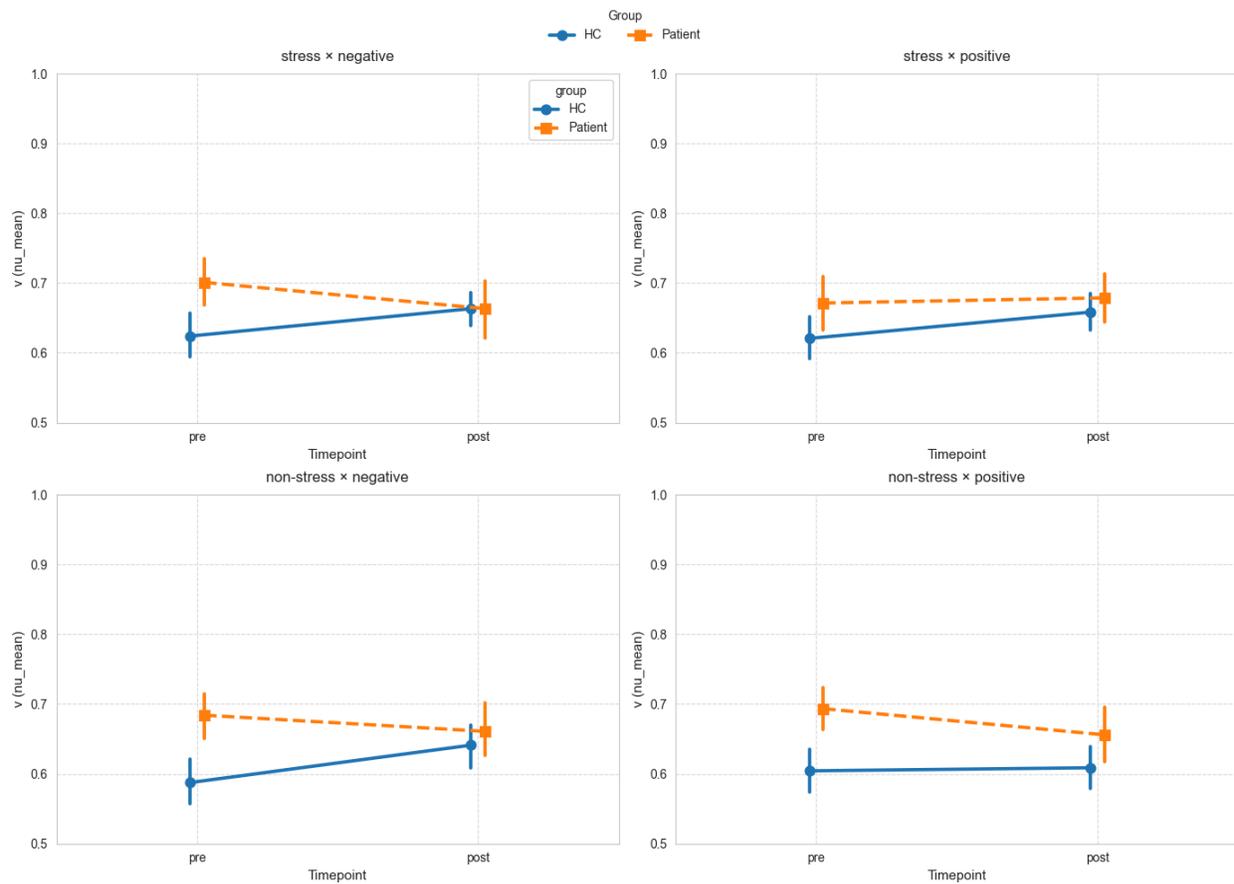

Figure S14: Change in v mean from pre to post manipulation by group and manipulation word valence combinations

*IMS items*

Participants are asked to rate how they feel on a 5-point scale for the items below.

1. Depressed-Happy
2. Distracted-Focused
3. Worthless-Valuable
4. Sleepy-Alert
5. Pessimistic-Optimistic
6. Apathetic-Motivated
7. Guilty-Proud
8. Numb-Interested
9. Withdrawn-Welcoming
10. Hopeless-Hopeful
11. Tense-Relaced
12. Worried-Untroubled

13. Fearful-Fearless
14. Anxious-Peaceful
15. Restless-Calm

*IMS tables*

*Overall IMS analysis*

We found that the main effect of normalized IMS rating change and the interaction between group and normalized IMS rating change was driven primarily by the non-stress runs. Looking at normalized change in IMS scores for non-stress manipulation runs and group as predictors on CH rates, we found a significant main effect of normalized IMS rating change ($b$ = 0.060, $SE$ = 0.022, $z$ = 2.750, $p$ = 0.006), a significant interaction between normalized IMS rating change and group ($b$ = -0.125, $SE$ = 0.041, $z$ = -3.053, $p$ = 0.002), but not a main effect of group ($b$ = 0.61, $SE$ = 0.045, $z$ = 1.366, $p$ = 0.172). Comparatively, when looking at the stress conditions, we did not see a significant main effect of normalized IMS rating change ($b$ = 0.020, $SE$ = 0.031, $z$ = 0.659, $p$ = 0.510), a significant interaction between normalized IMS rating change and group ($b$ = -0.055, $SE$ = 0.053, $z$ = -1.037, $p$ = 0.300), nor a significant main effect of group ($b$ = 0.094, $SE$ = 0.064, $z$ = 1.455, $p$ = 0.146). For v, we did not see an overall main effect nor interaction between normalized IMS rating change and group across all run types. However, we did find a significant interaction between normalized IMS rating change and group across non-stress runs ($b$ = -0.149, $SE$ = 0.065, $z$ = -2.278, $p$ = 0.023), when v increased, normalized IMS ratings went down, but there was no main effect of group nor normalized IMS rating change.

Table S15: Change in IMS scores by group

| group | Mean change | Std change | Mean absolute change | Std absolute change | Min change | Max change | Baseline mean | Mean rating |
|---|---|---|---|---|---|---|---|---|
| HC | 9.300 | 20.617 | 14.440 | 17.373 | -25.0 | 75.0 | 80.760 | 71.460 |
| Patient | 2.684 | 13.994 | 9.211 | 10.825 | -31.0 | 45.0 | 62.882 | 60.197 |

Table S16: Change in IMS scores by group and manipulation valence

| group | Manipulation Valence | Mean change | Mean absolute change | Std change | Min change | Max change | Rating mean |
|---|---|---|---|---|---|---|---|
| HC | stress | 10.143 | 15.000 | 21.914 | -25.0 | 74.0 | 70.776 |
| HC | non-stress | 8.490 | 13.902 | 19.475 | -23.0 | 75.0 | 72.118 |
| Patient | stress | 2.512 | 9.439 | 14.591 | -31.0 | 45.0 | 60.268 |

| | | | | | | |
|---|---|---|---|---|---|---|
| Patient | non-stress | 2.886 | 8.943 | 13.471 | -21.0 | 44.0 | 60.114 |

*Analysis of fatigue IMS item*

We looked at the effects of the manipulations overall on the fatigue item of the IMS (Sleepy-Alert). First, we looked at the overall changes in the fatigue item as a function of group. We a significant difference in change in fatigue from baseline over all groups and run types ($b = -0.576$, $SE = 0.136$, $z = -4.239$, $p < 0.001$) where mean fatigue was significantly higher post manipulation compared to baseline (measured at the beginning of the first run). We did not see any significant main effect of group on baseline fatigue ($b = -0.179$, $SE = 0.202$, $z = -0.886$, $p = 0.376$), nor change in fatigue as a predictor of group ($b = -0.115$, $SE = 0.484$, $z = -0.238$, $p = 0.812$). Next we looked at manipulation valence and group as a predictor for change in fatigue post manipulation and did not find any significant interaction between group and manipulation valence (non-stress runs compared to baseline: $b = 0.111$, $SE = 0.484$, $z = 0.229$, $p = 0.819$; stress runs compared to baseline: $b = 0.050$, $SE = 0.484$, $z = 0.103$, $p = 0.918$). We also looked at how the change in fatigue and group predicted CH rates and v values, but did not find any significant main effects (CH rate ~ fatigue change: $b = 0.002$, $SE = 0.014$, $z = 0.106$, $p = 0.915$; v mean ~ fatigue change: $b = -0.020$, $SE = 0.023$, $z = -0.873$, $p = 0.383$), nor interactions between group and change in fatigue as predictors for CH rates and v values (CH rate ~ group * fatigue change: $b = 0.015$, $SE = 0.018$, $z = 0.850$, $p = 0.395$; v mean ~ group * fatigue change: $b = 0.021$, $SE = 0.028$, $z = 0.764$, $p = 0.445$).

Table S17: Change in fatigue item from the IMS from baseline compared to post manipulation by manipulation valence and group

| group | Manipulation Valence | Mean fatigue change from baseline | Std fatigue change from baseline |
|---|---|---|---|
| HC | stress | -0.987 | 1.575 |
| HC | non-stress | -1.144 | 1.376 |
| Patient | stress | -0.900 | 2.249 |
| Patient | non-stress | –1.032 | 2.125 |

Table S18: Change in fatigue item from the IMS from baseline compared to post manipulation by manipulation-word valence combinations and group

| group | Manipulation Valence | Word Valence | Mean fatigue change from baseline | Std fatigue change from baseline |
|---|---|---|---|---|
| HC | stress | Negative | -1.325 | 1.749 |

| | | | Positive | -0.611 | 1.301 |
|---|---|---|---|---|---|
| | HC | non-stress | Negative | -1.000 | 1.490 |
| | | | Positive | -1.272 | 1.294 |
| Patient | | stress | Negative | -1.219 | 2.366 |
| | | | Positive | –0.536 | 2.135 |
| Patient | | non-stress | Negative | -1.603 | 2.366 |
| | | | Positive | -1.000 | 1.918 |

*Analysis of attention IMS item*

Looking specifically at the attention item from the IMS, we wanted to see if the type of manipulation changed how attentive participants were after the manipulation. Overall for the attention item, we did not see an effect of group on the change in attention before and after the manipulation ($b = 0.273$, $SE = 0.597$, $z = 0.456$, $p = 0.648$), nor in mean attention across runs ($b = -0.569$, $SE = 0.556$, $z = -1.024$, $p = 0.306$). However, when we looked group, manipulation valence, and session type as predictors for mean attention, we found a significant three way interaction where patients in the stress AVAM condition showed lower attention scores compared ($b = -1.363$, $SE = 0.684$, $z = -1.992$, $p = 0.046$). When we looked specifically in the patient group with manipulation valence and session type as predictors for mean attention, we found a significant main effect of session type ($b = 0.898$, $SE = 0.436$, $z = 2.062$, $p = 0.039$) where mean attention scores in the AVAM sessions were lower than MIST sessions and we also found a significant interaction between manipulation valence and session type ($b = -1.170$, $SE = 0.525$, $z = -2.229$, $p = 0.026$) where mean attention scores were lower in the AVAM sessions in the stress conditions.

Tables S19: Change and mean fatigue in attention item from the IMS from baseline compared to post manipulation by manipulation valence, group, and session type (AVAM/MIST)

| group | Manipulation Valence | Session type | Mean attention scores | Std attention scores | Mean attention change from baseline | Std attention change from baseline |
|---|---|---|---|---|---|---|
| HC | stress | AVAM | 4.778 | 2.067 | -0.583 | 1.849 |
| | | MIST | 3.775 | 1.400 | -1.200 | 1.735 |
| HC | non-stress | AVAM | 4.272 | 1.734 | -0.500 | 1.611 |
| | | MIST | 4.912 | 1.079 | -0.781 | 1.629 |
| Patient | stress | AVAM | 3.333 | 1.759 | -1.067 | 1.999 |
| | | MIST | 4.867 | 1.457 | 0.533 | 1.274 |
| Patient | non-stress | AVAM | 3.750 | 1.198 | -0.375 | 1.737 |
| | | MIST | 4.038 | 1.282 | -0.423 | 1.305 |

## *VAS tables*

For the affective slider, participants are asked to rate how they feel from sleepy-alert and unpleasant-pleasant.

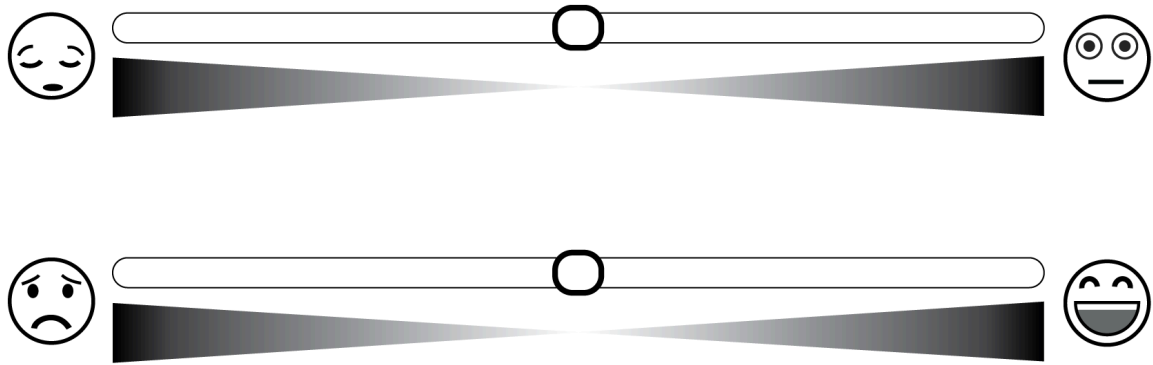

Figure S15: Affective slider. We used a previously validated affective slider scale to measure pleasure and arousal at four points during each run (Betella & Verschure, 2016) . Participants were instructed to rate how they felt by moving the circle along the slider.

### *Arousal scores decreased as CH rates increased in the patient group, but not in the control group*

We calculated the mean change of arousal scores by taking the mean score of the arousal slider administered before the manipulation and subtracting it from the arousal slider administered after the manipulation. When looking at how normalized change in arousal scores and group predicted CH rates across all run types using a mixed linear model, we did not find significant group differences ($b$ = 0.062, $SE$ = 0.053, $z$ = 1.173, $p$ = 0.241) nor did we find an effect of arousal ($b$ = -0.029, $SE$ = 0.018, $z$ = -1.583, $p$ = 0.113), but we did find a significant interaction between group and arousal difference ($b$ = 0.053, $SE$ = 0.021, $z$ = 2.465, $p$ = 0.014). We found that as arousal scores increased in the patient group, CH rates went down and vice versa in the control group.

We found that the interaction between normalized arousal scores and group as predictors for CH rates was mostly driven by non-stress run types. We found a significant interaction between normalized change in arousal scores and group as predictors for CH rates across non-stress runs ($b$ = 0.096, $SE$ = 0.034, $z$ = 2.824, $p$ = 0.005), but no main effect of group ($b$ = 0.053, $SE$ = 0.052, $z$ = 1.015, $p$ = 0.310) nor a main effect of normalized arousal difference ($b$ = -0.055, $SE$ =

0.030, $z = -1.853$, $p = 0.064$). We did not find a significant interaction between normalized change in arousal scores and group as predictors for CH rates in the stress condition ($b = 0.002$, $SE = 0.030$, $z = 0.060$, $p = 0.952$). Similar to what we saw above when grouping across all run types, as arousal scores went up in the patient group, CH rates went down and vice versa in the control group. We did not see any significant group differences, arousal differences, nor interactions between group and arousal differences for the stress manipulation run types. For the pleasure metric, we did not see any significant group differences, pleasure differences in different manipulation valence conditions, not any interactions between group and pleasure differences in different manipulation valence conditions.

Table 20: Change in arousal and pleasure scores (baseline-post manipulation) by group

| group | baseline_arousal | mean_arousal_diff | std_arousal_diff | baseline_pleasure | mean_pleasure_diff | std_pleasure_diff |
|---|---|---|---|---|---|---|
| HC | 0.515 | -0.072 | 0.165 | 0.549 | 0.027 | 0.287 |
| Patient | 0.553 | -0.048 | 0.290 | 0.481 | -0.021 | 0.253 |

Table 21: CH rates as a function of change in arousal and pleasure scores (baseline-post manipulation) by group and manipulation valence

| group | manipulation Valence | mean_arousal_diff | std_arousal_diff | mean_pleasure_diff | std_pleasure_diff |
|---|---|---|---|---|---|
| HC | stress | -0.070 | 0.184 | 0.110 | 0.299 |
| HC | non-stress | -0.072 | 0.157 | -0.072 | 0.261 |
| Patient | stress | 0.013 | 0.301 | 0.042 | 0.251 |
| Patient | non-stress | -0.076 | 0.278 | -0.055 | 0.258 |

*AVAM Validation*

To validate the audiovisual affective manipulation (AVAM), participants from MTURK (n = 82) were assigned to one of six possible manipulations: three stress and three non-stress (these included the same content, and the same music, but different orderings of the images). Measurements of valence and arousal were performed before and after the manipulation.

Table S22: Mean valence ratings for stress and non-stress manipulations

| | n | Mean (μ) | Test statistic | df | p-value |
|---|---|---|---|---|---|
| Stress | 41 | 4.99 | 6.5691 | 3.3655 | 0.004982 |

| | | |
|---|---|---|
| Non-stress | 41 | 1.63 |

All three stress manipulations resulted in a mean decrease in valence across participants (μ = -4.53, -5.86, -4.58), and all three non-stress manipulations resulted in a mean increase (μ = +2.08, +1.67, +1.14). Analysis using a Welch Two Sample T-test showed significant differences (p < 0.005) in the valence measurements taken before and after the manipulation.

Table S23: Mean arousal ratings for stress and non-stress manipulations

| | n | Mean (μ) | Test statistic | df | p-value |
|---|---|---|---|---|---|
| Stress | 41 | 3.80 | 10.47 | 2.4558 | 0.004254 |
| Non-stress | 41 | 0.28 | | | |

All three stress manipulations resulted in a mean decrease in arousal across participants (μ = -3.20, -4.29, -3.92). While all three non-stress manipulations increased mean arousal, the increase was much lower (μ = +0.33, +0.07, +0.43) compared to the relative decrease due to negative stimuli. Analysis using a Welch Two Sample T-test showed significant differences (p < 0.005) in the arousal measurements taken before and after the manipulation.

These results support the use of the AVAM in using manipulations to elicit affective responses in participants. Stress manipulations successfully induced negative affective states (lowered valence and arousal), and non-stress manipulations induced positive affective states (increased valence and arousal).

### AHRS and LSHS correlations with CH rates and v across sessions

| | All | Patient | Control |
|---|---|---|---|
| Mean AHRS scores | Mean: 8.222<br>Std: 10.453 | Mean: 17.750<br>Std: 8.551 | Mean: 0.600<br>Std: 2.324 |

Mean AHRS scores at baseline were significantly higher in the patient compared to control groups ($U = 2.0$, $p < 0.001$). Within the patient group, patients with FEP compared to CHR-P patients had higher AHRS scores, although this difference (AHRS: $U = 29.5$, $p = 0.061$) was not significant.

Counterintuitively given the LSHS results and previous work, there was a significant negative correlation between AHRS scores and v values for the patient group in session 1 ($\rho$ = –0.65, $p$ = .023; see figure S16). Looking specifically at the frequency of hallucinations item on the AHRS (AHRS1), we did not find any significant correlation between AHRS1 and CH rates (session 1 HC and patients: $\rho$ = 0.07, $p$ = .736; session 1 patients only: $\rho$ = -0.24, $p$ = .453; session 2 HC and patients: $\rho$ = 0.37, $p$ = .142; session 2 patients only: $\rho$ = -0.15, $p$ = .749), however, with v values, we found a strong negative correlation between AHRS1 and v values in the patient group for session 1 while all other correlations were non-significant (session 1 HC and patients: $\rho$ = -0.05, $p$ = .796; session 1 patients only: $\rho$ = -0.68, $p$ = .016; session 2 HC and patients: $\rho$ = 0.14, $p$ = .601; session 2 patients only: $\rho$ = -0.13, $p$ = .780. Looking specifically at the distress of hallucinations item on the AHRS (AHRS7), we did not find any significant correlation between AHRS7 and CH rates (session 1 HC and patients: $\rho$ = 0.11, $p$ = .569; session 1 patients only: $\rho$ = -0.18, $p$ = .566; session 2 HC and patients: $\rho$ = 0.05, $p$ = .842; session 2 patients only: $\rho$ = -0.49, $p$ = .268) nor v values (session 1 HC and patients: $\rho$ = 0.19, $p$ = .356; session 1 patients only: $\rho$ = -0.30, $p$ = .343; session 2 HC and patients: $\rho$ = 0.12, $p$ = .651; session 2 patients only: $\rho$ = -0.14, $p$ = .761) for patients and controls together, as well as in patients only. See Table SX for the full breakdown of correlations between symptom scores on CH rates and v values for sessions 1 and 2. For AHRS scores, 40% or more of patients had not had recent hallucinations, and most had not had frequent ones. Therefore, we determined that the AHRS was not particularly valid in this small population.

Table S24: Correlation between AHRS/LSHS scores by CH rates/v for both groups combined and patients only

| Correlation | Groups | Session | Spearman Coefficient | p-value |
|---|---|---|---|---|
| AHRS vs CH rates | All | 1 | 0.05 | 0.814 |
| AHRS vs CH rates | Patient | 1 | -0.16 | 0.593 |
| AHRS vs v values | All | 1 | 0.12 | 0.552 |
| AHRS vs v values | Patient | 1 | -0.65 | 0.023 |
| AHRS vs CH rates | All | 2 | 0.40 | 0.111 |
| AHRS vs CH rates | Patient | 2 | -0.29 | 0.531 |
| AHRS vs v values | All | 2 | 0.15 | 0.574 |

| AHRS vs v values | Patient | 2 | -0.56 | 0.193 |
| --- | --- | --- | --- | --- |
| LSHS vs CH rates | All | 1 | 0.42 | 0.030 |
| LSHS vs v values | Patient | 1 | 0.43 | 0.165 |
| LSHS vs CH rates | All | 1 | 0.47 | 0.014 |
| LSHS vs v values | Patient | 1 | 0.44 | 0.152 |
| LSHS vs CH rates | All | 2 | 0.34 | 0.181 |
| LSHS vs v values | Patient | 2 | 0.22 | 0.631 |
| LSHS vs CH rates | All | 2 | 0.18 | 0.494 |
| LSHS vs v values | Patient | 2 | -0.21 | 0.658 |

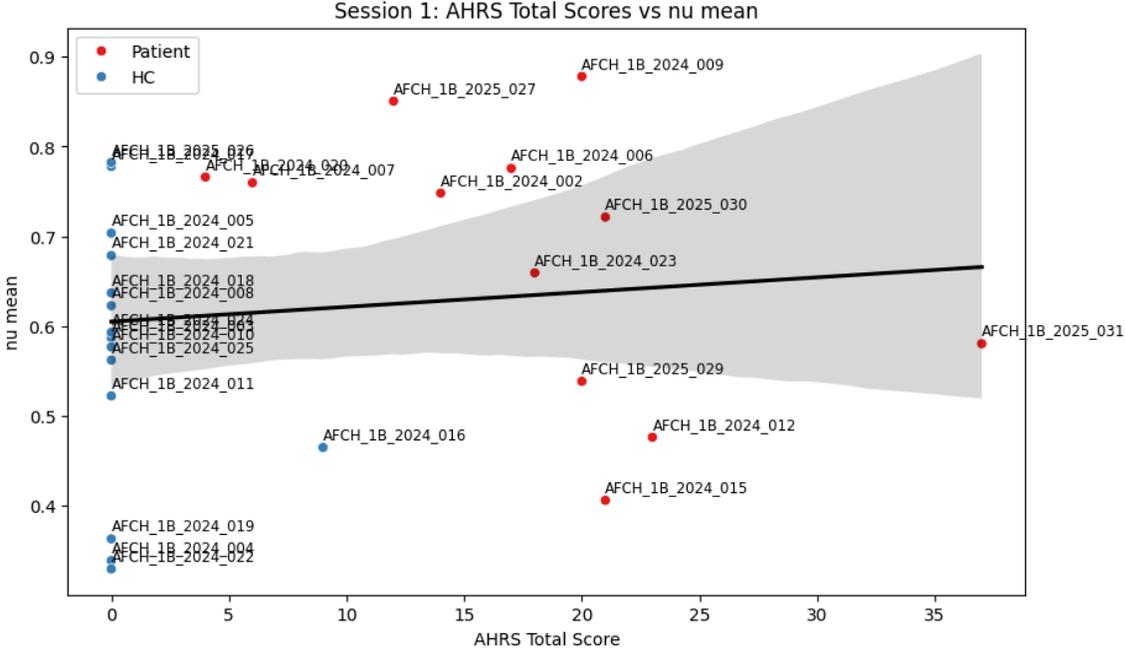

Figure S16: Session 1 correlation of v means and AHRS scores:
Session 1 Spearman Correlation (All): 0.12, p-value: 0.552
Session 1 Spearman Correlation (Patients only): -0.65, p-value: 0.023

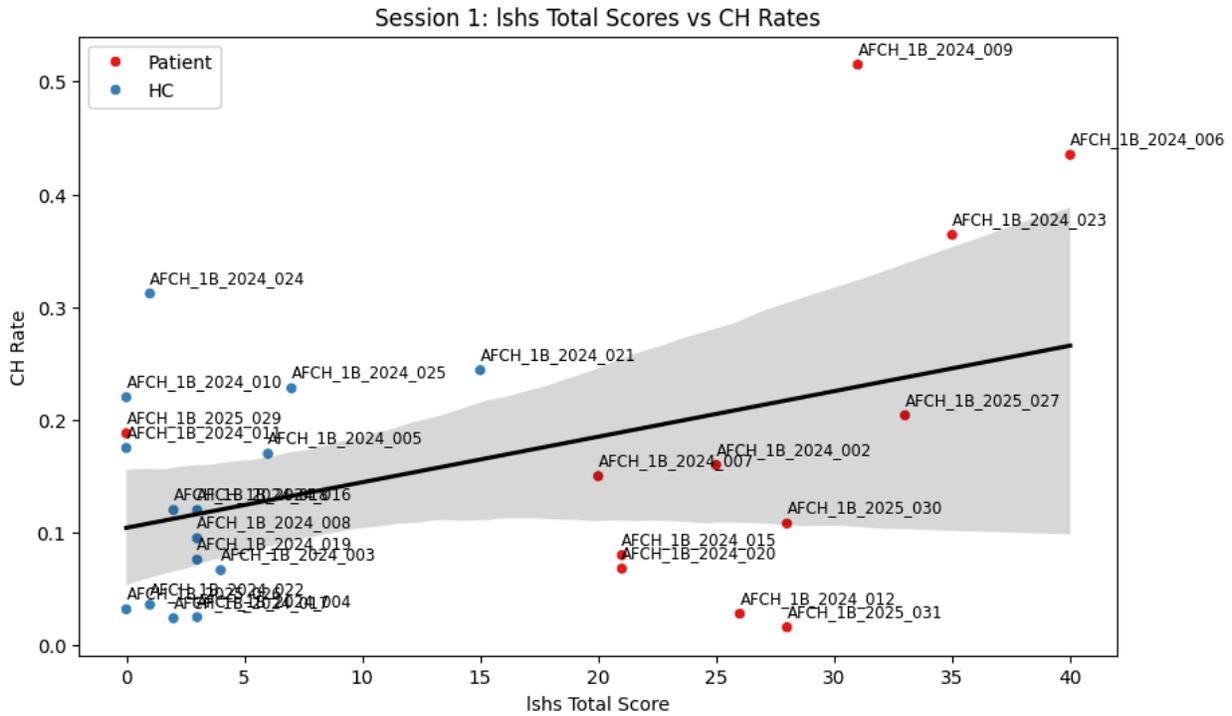

Figure S17: Session 1 correlation of CH rates and LSHS scores:
Session 1 Spearman Correlation (All): 0.42, p-value: 0.030
Session 1 Spearman Correlation (Patients only): 0.43, p-value: 0.165

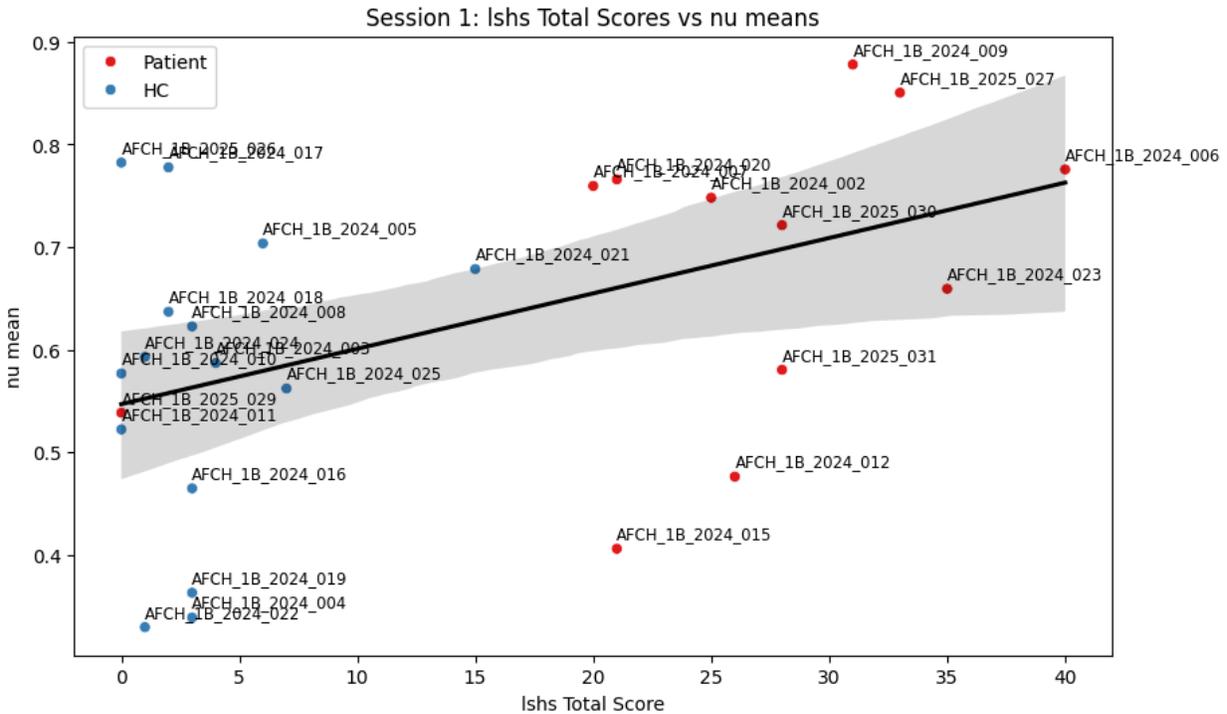

Figure S18: Session 1 correlation of v means and LSHS scores:
Session 1 Spearman Correlation (All): 0.47, p-value: 0.014
Session 1 Spearman Correlation (Patients only): 0.44, p-value: 0.152

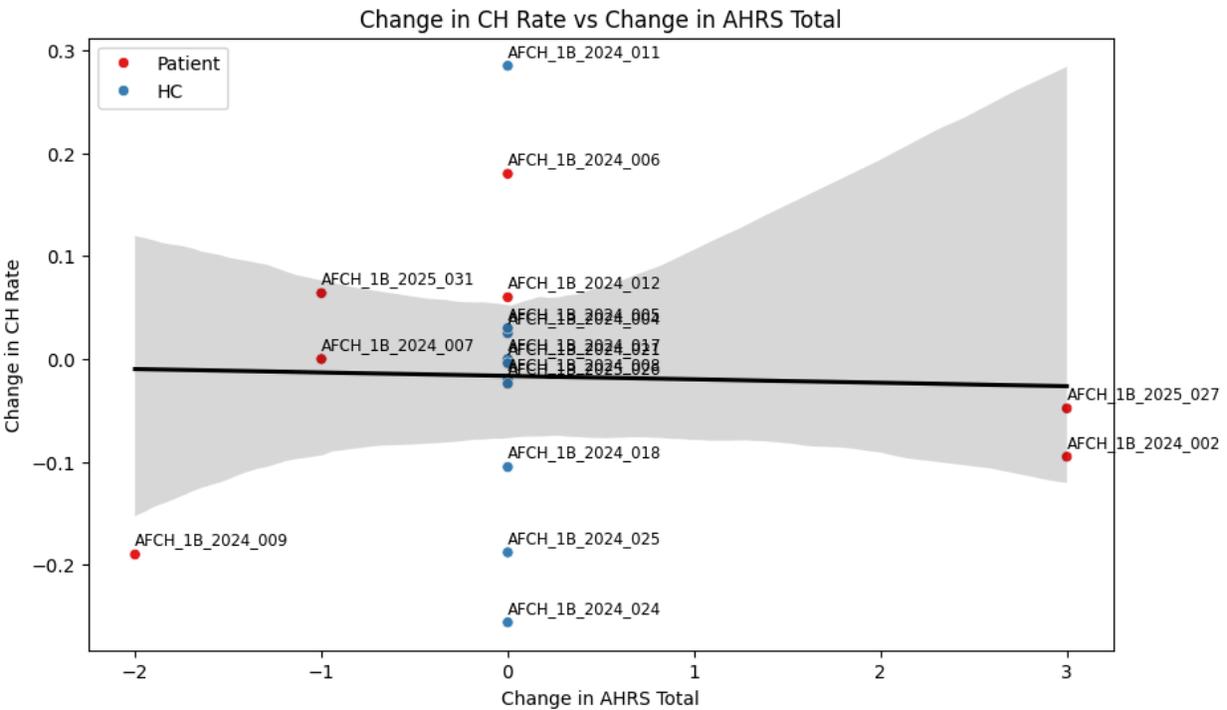

Figure S19: Correlation between change in CH rates and change in AHRS scores across sessions

Changes in AHRS Correlation (All): -0.13, p-value: 0.6243
Changes in AHRS Correlation (Patients only): 0.02, p-value: 0.9688

![Change in CH Rate vs Change in LSHS Total scatter plot]

Figure S20: Correlation between change in CH rates and change in LSHS scores across sessions
Changes in LSHS Correlation (All): -0.03, p-value: 0.9001
Changes in LSHS Correlation (Patients only): -0.40, p-value: 0.3786

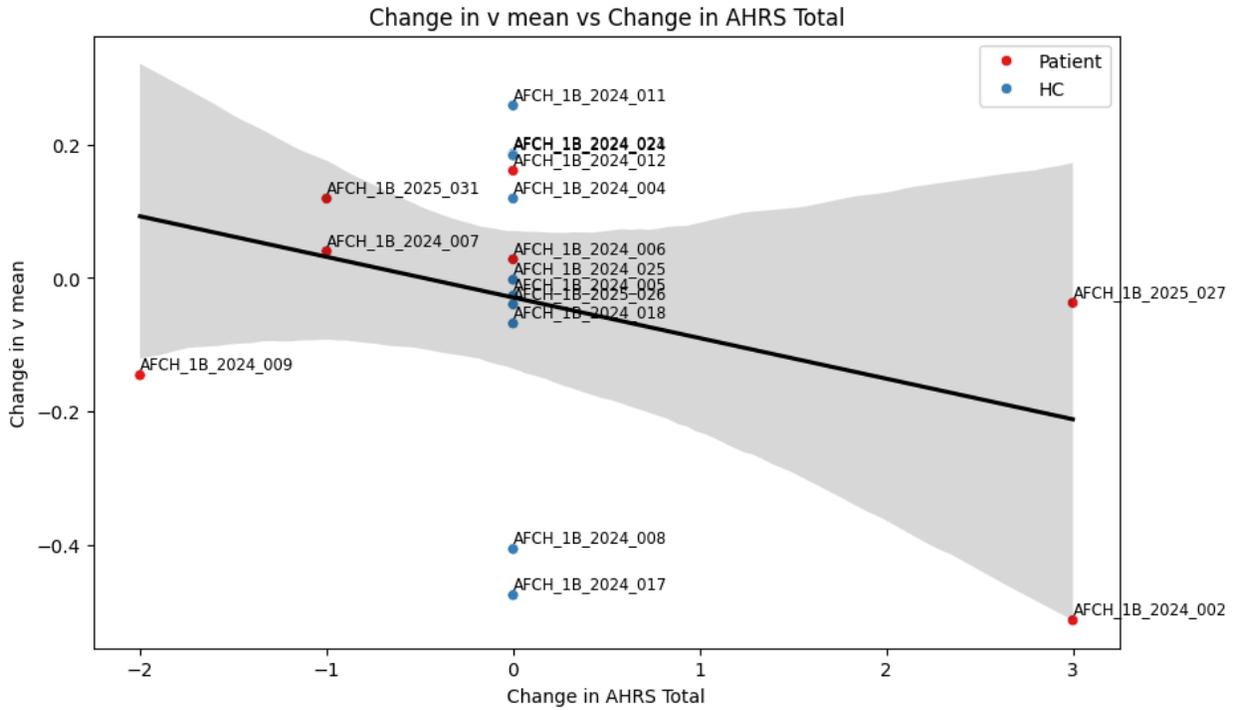

Figure S21: Correlation between change in v values and change in AHRS scores across sessions
Changes in AHRS Correlation (All): -0.19, p-value: 0.4641
Changes in AHRS Correlation (Patients only): -0.26, p-value: 0.5780

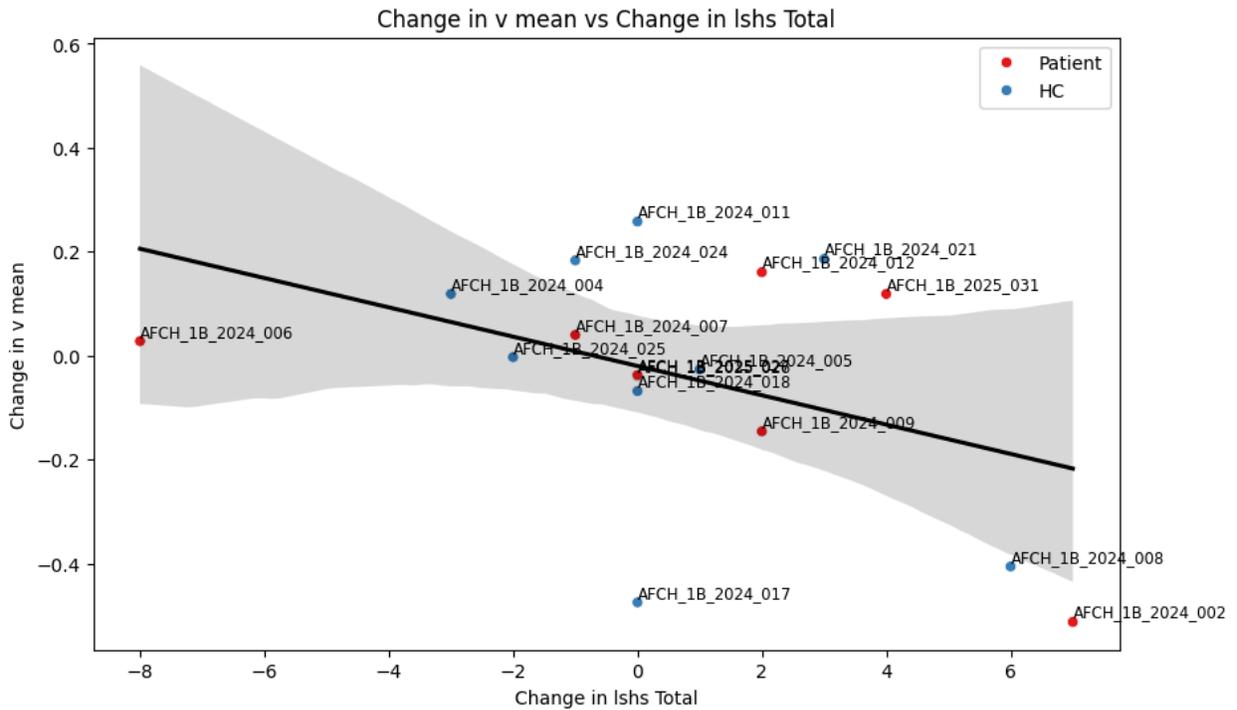

Figure S22: Correlation between change in v values and change in LSHS scores across sessions
Changes in lshs Correlation (All): -0.30, p-value: 0.2492
Changes in lshs Correlation (Patients only): -0.20, p-value: 0.6701

*Word Validation*

To validate the words used in the experiment, measurements of valence, arousal, and dominance were performed in association with each word used. There are six negative, two neutral, and six positive words for a total of fourteen words.

Table S25: Mean valence, arousal, and dominance measurements for each negative word.

| - | Valence | | Arousal | | Dominance | |
|---|---|---|---|---|---|---|
| Word | Mean | SD | Mean | SD | Mean | SD |
| Bad | 4.057692 | 2.042811 | 6.423077 | 1.818706 | 4.826923 | 2.272781 |
| Die | 3.857143 | 2.625025 | 4.742857 | 3.003359 | 3.457143 | 2.512824 |
| Dumb | 4.205882 | 2.446759 | 5.382353 | 2.322677 | 4.500000 | 2.351660 |
| Hate | 4.063830 | 1.904277 | 5.382979 | 2.373341 | 4.446809 | 2.030298 |
| Kill | 3.171429 | 2.490992 | 3.857143 | 2.702318 | 3.600000 | 2.534468 |
| Stress | 4.317073 | 2.229787 | 5.292683 | 2.512408 | 4.682927 | 2.296073 |
| **ALL** | 3.945508 | 0.2795436 | 5.180182 | 0.3985489 | 4.252300 | 0.1843328 |

Table S26: Mean valence, arousal, and dominance measurements for each positive word.

| + | Valence | | Arousal | | Dominance | |
|---|---|---|---|---|---|---|
| Word | Mean | SD | Mean | SD | Mean | SD |
| Fun | 6.729730 | 2.103551 | 6.162162 | 2.432574 | 6.162162 | 2.230016 |
| Good | 7.170732 | 1.579596 | 6.875000 | 2.114510 | 7.051282 | 1.700599 |
| Joy | 7.485714 | 1.541035 | 6.314286 | 2.323428 | 6.714286 | 1.840259 |
| Hug | 6.659091 | 1.975939 | 7.1395535 | 1.934418 | 6.441860 | 2.249892 |
| Safe | 6.571429 | 1.632135 | 6.823529 | 2.007117 | 6.735294 | 2.004674 |
| Kind | 7.000000 | 1.748015 | 7.513514 | 1.709802 | 6.837838 | 1.907784 |
| **ALL** | 6.936116 | 0.2287236 | 6.804671 | 0.2638248 | 6.657120 | 0.2183100 |

A three-way ANOVA was used to compare the positive, neutral, and negative means for valence and arousal, respectively. A statistically significant difference between groups for both valence (Table S27) and arousal (Table S28) was found between the three word groups. Pairwise comparisons between word groups revealed statistically significant differences in valence between all three groups (Table S29). Pairwise comparisons revealed significant differences in arousal between only positive and negative words (Table S29).

Table S27: results of a three-way ANOVA comparing the effect of positive, negative, and neutral means for valence.

| DF | Sum of squares | Mean squares | F value | p value |
|---|---|---|---|---|
| Category | 2 | 26.997 | 13.498 | 0.0000001 |
| Residuals | 11 | 1.453 | 0.132 | NA |

Table S28: results of a three-way ANOVA comparing the effect of positive, negative, and neutral means for arousal.

| DF | Sum of squares | Mean squares | F value | p value |
|---|---|---|---|---|
| Category | 2 | 9.449 | 4.725 | 0.0027421 |
| Residuals | 11 | 4.914 | 0.441 | NA |

Table S29: pairwise comparison between positive and negative words for valence and arousal

|  | Valence p-value | Arousal p-value |
|---|---|---|
| Positive-Negative | 0.0000001 | 0.0038272 |

These results indicate that there are differences in participant valence between the positive, neutral, and negative words, and that there are differences in participant arousal between the positive and negative words. Mean valence is significantly lower in response to negative words than to neutral or positive words; mean arousal is significantly lower in response to negative words than to positive words. This data reflects appropriate use of these words in the MIST to generate affective responses.

**Confidence ratings**

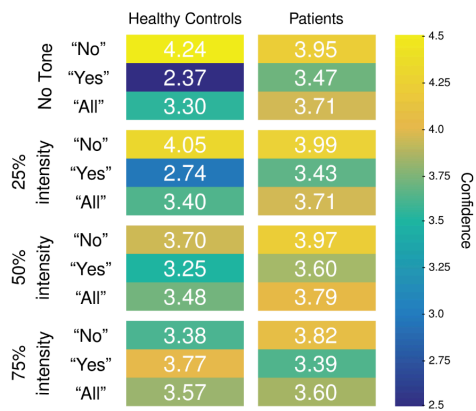
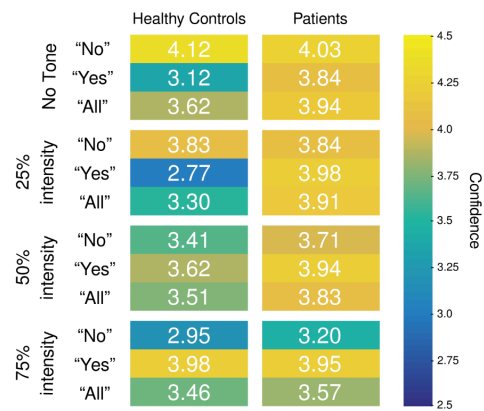

Confidence ratings under stress manipulation and negative word valence     Confidence ratings under stress manipulation and positive word valence

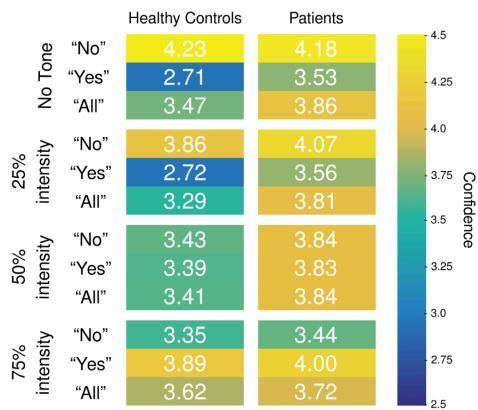
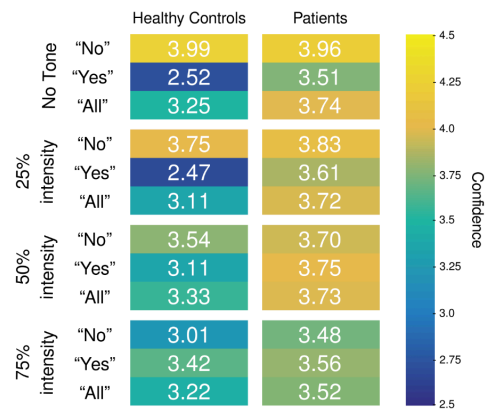

Confidence ratings under non-stress manipulation and negative word valence     Confidence ratings under non-stress manipulation and positive word valence

Figure S23: Confidence ratings by Manipulation-Word valence combinations

Linear mixed-effects models were conducted to examine the effect of manipulation-word valence combinations on confidence ratings. The model included valence combination as a fixed effect and random intercepts for participants to account for within-subject variability. The results revealed a significant main effect of valence combinations on confidence ratings. We found significantly higher confidence ratings in the stress-negative conditions compared to the non-stress-positive conditions in almost all of the trial and response types(No word$_{\text{all responses}}$: *b = -0.141, SE = 0.033, z = -4.209, p < 0.001*; 25%$_{\text{all responses}}$ : *b = -0.206, SE = 0.054, z = -3.809, p < 0.001*; 25%$_{\text{"yes"}}$ : *b = -0.214, SE = 0.094 z = -2.281, p = 0.023*; 50%$_{\text{all responses}}$ : *b = -0.124, SE = 0.050, z = -2.468, p = 0.014*; 50%$_{\text{"yes"}}$ : *b = -0.170, SE = 0.065, z = -2.637, p = 0.008*;75%$_{\text{all responses}}$ : *b = -0.225, SE = 0.042, z = -5.370, p < 0.001*; 75%$_{\text{"yes"}}$ : *b = -0.189, SE = 0.042, z = -4.466, p < 0.001*; , the difference was not significant in the CH trials but the numerical value showed the same trend. We also found that the stress-negative had lower ratings than the nonstress-negative when there was indication of hearing a sound in the 75% intensity trial (*b = 0.270, SE = 0.042, z = 6.476, p < 0.001*) and across all responses for the no word and the 75% intensity trials (No word: *b = 0.108, SE = 0.034, z = 3.203, p = 0.001; 75%: b = 0.211, SE =*

0.042, $z$ = 5.012, $p < 0.001$). Additionally, stress-negative confidence ratings were lower than stress-positive for all trials with indication of hearing a sound(No Tone: $b$ = 0.288, $SE$ = 0.091, $z$ = 3.151, $p = 0.002$; 50%: $b$ = 0.162, $SE$ = 0.063, $z$ = 2.573, $p = 0.010$; 75%: $b$ = 0.263, $SE$ = 0.042, $z$ = 6.244, $p < 0.001$), except for the 25% intensity where stress-negative condition was higher than stress-positive across all responses ($b$ = -0.111, $SE$ = 0.054, $z$ = -2.060, $p = 0.039)$. The model included a random intercept variance of around 1 across trial types, indicating a substantial variability in overall rating levels in participants.